\documentclass[11pt,a4paper]{article}
\pdfoutput=1 
\usepackage{jheppub2}
\usepackage[table]{xcolor}
\usepackage[utf8]{inputenc}
\usepackage{colortbl}
\usepackage{varwidth}
\RequirePackage{ifpdf} 
\usepackage{amsmath} 
\usepackage{mathtools}
\usepackage{pstricks}
\usepackage[final]{pdfpages} 
\usepackage{ifpdf} 
\usepackage{slashed}
\usepackage{natbib}
\setcitestyle{square, comma, numbers,sort&compress}
\usepackage{hyperref}
\usepackage[normalem]{ulem}
\usepackage{color}
\usepackage{soul,xcolor}
\usepackage{xcolor}
\definecolor{urlblue}{rgb}{0.2,0.4,0.7}
\definecolor{citegreen}{rgb}{0,0.6,0.2}
\definecolor{linkred}{rgb}{0.9,0.2,0.1}
\setstcolor{red}
\usepackage{hyperref}
\hypersetup{
colorlinks=true, citecolor=citegreen, linkcolor=blue, urlcolor=urlblue}
\usepackage{float}
\usepackage[caption = false]{subfig}
\usepackage{graphics}
\usepackage{etoolbox} 
\usepackage{fixmath}
\usepackage{psfrag}
\usepackage[utf8]{inputenc} 
\usepackage[T1]{fontenc}

\usepackage{amsfonts}
\usepackage{autobreak}
\usepackage{marginnote}
\usepackage{enumitem}
\usepackage{appendix}

\newcommand{\NOdisplay}[1]{ }
\DeclareUnicodeCharacter{00A0}{ }
\DeclareUnicodeCharacter{03B2}{\ensuremath{\beta}}
\DeclareUnicodeCharacter{03B5}{ }
\DeclareUnicodeCharacter{203E}{ }
\DeclareUnicodeCharacter{03D5}{ }

\newcommand{\iu}{{i\mkern1mu}}
\DeclareMathOperator{\Li}{Li}

\title{Master integrals for $\mathcal{O}(\alpha \alpha_s)$ corrections to $H \rightarrow Z Z^*$}

\author{Ekta Chaubey$^{a}$}
\emailAdd{ekta@to.infn.it}
\affiliation{
$^a$Dipartimento di Fisica and Arnold-Regge Center, Universit\`a di Torino, 
\\ and INFN, Sezione di Torino, Via Pietro Giuria 1, I-10125 Torino, Italy}

\author{Mandeep Kaur$^{b}$}
\emailAdd{mandeepkaur.iiser@gmail.com}
\affiliation{
$^b${Indian Institute of Science Education and Research Mohali, \\Knowledge City, Sector 81, SAS Nagar, Punjab 140306, India}}

\author{Ambresh Shivaji$^{b}$}
\emailAdd{ashivaji@iisermohali.ac.in}

\preprint{}

\abstract{We present analytic results for all the Feynman integrals relevant for ${\mathcal O}(\alpha \alpha_s)$ virtual corrections to $H \rightarrow ZZ^*$ decay. We use the method of differential equations to solve the master integrals while keeping the full dependence on the masses of all the particles, including those propagating in the loop. Due to the presence of four mass scales, we encounter multiple square roots. We argue that all the occurring square roots can not be rationalized at the same time as a simultaneous rationalization brings us to integrals over $CY_3$ manifolds. Hence we rationalize only three square roots simultaneously and construct suitable ansätze to obtain dlog-forms containing the square roots after obtaining an epsilon-factorised form for the differential equations in $D=4-2\epsilon$ dimensions. We present the alphabet and the analytic form of all the boundary constants that appear in the solutions of the differential equations. The results for master integrals are expressed in terms of Chen's iterated integrals with dlog one-forms.  
}

\begin{document}
\allowdisplaybreaks[4]
\unitlength1cm
\maketitle
\flushbottom

\section{Introduction}
\label{sec:intro}

Ever since the Higgs boson discovery~\citep{CMS:2012qbp, ATLAS:2012yve}, the particle has always remained in the spotlight. It is believed that more profound studies of Higgs boson properties can shed some light on the mysteries of nature and can also help to probe the beyond Standard Model (BSM) physics. 
In the light of no clear signature of new physics emerging from the LHC data analysed so far, precision measurements in the Higgs sector 
are one of the main goals of future colliders such as FCC-hh \citep{Benedikt:2022kan}.  
A better understanding of Higgs properties requires precise theoretical predictions for Higgs production and decay channels. 
Out of five prominent decay modes of the Higgs, $H \rightarrow ZZ^*$ decay, where $Z^*$ is off-shell, is a rare one. When the $Z$ and $Z^*$ further decay into a pair of charged leptons, it is also known as the ``Golden decay channel''. Due to the 
properties like a clean background and excellent resolution in reconstructing Higgs, this decay channel played a very important role in the Higgs discovery. The decay channel, due to its non-trivial final state kinematics,
has also been utilized to study the spin and $CP$ properties of the Higgs boson \citep{Bolognesi:2012mm, CMS:2014nkk, ATLAS:2013xga}. 
On the standard model side, it is crucial to have precise predictions including the higher-order terms in the perturbative 
calculation of the amplitude for the partial decay width of the $H \to ZZ^*$. 

The complete Next-to-leading (NLO) electroweak corrections to $H \to 4 f$ for semi-leptonic, leptonic and hadronic final states have been calculated by Bredenstein, Denner, Dittmaier, and Weber \citep{Bredenstein_2007, BREDENSTEIN2006131} and calculations are encoded in a publicly available Monte Carlo (MC) code {\tt Prophecy4f} \citep{Bredenstein:2007ec}. Using the code, NLO QCD corrections in semi-leptonic and hadronic channels can also be obtained. 
The NLO electroweak corrections matched with QED Parton Shower (PS) for the Higgs decay into four charged leptons with off-shell $Z$-bosons have also been calculated. {The entire} calculation is implemented in a publicly available event generator {\tt Hto4l} \citep{Boselli:2015aha}. 
The QCD virtual corrections in decay modes with semi-leptonic ($H \to 2\ell 2q$) and hadronic ($H \to 4q$) final states appear at the one-loop level via correction to the $Zqq$ vertex. 
In the $H \to 4\ell$ decay mode with charged leptons in the final state, the QCD virtual corrections, however, involve $HZZ$ vertex corrections at the two-loop level with massive quarks in the loop. In perturbation theory, it is an ${\cal O}(\alpha \alpha_s)$ effect. The relevant {two-loop triangle} Feynman diagrams are similar to those which
appear in the QCD corrections to $H \to \gamma\gamma, \gamma Z$ decays. Due to the presence of many mass scales, the analytic calculation of two-loop { three-point} Feynman diagrams relevant to QCD corrections in $H \to 4\ell$ decay is quite non-trivial. 
Efforts are going on to provide these corrections by evaluating the two-loop Feynman integrals numerically \citep{Kaur:2022xxx}.

State-of-the-art for analytic studies of two-loop three-point functions including massive propagators are as follows.
For two-loop master integrals which appear in the virtual QCD corrections of $H \to \gamma\gamma$, the analytic results in terms of harmonic polylogarithms {have been} known for quite some time \citep{Aglietti:2006tp}.
Also, the analytic results of master integrals for massless 2-loop 3-point functions relevant for QCD corrections to decay $H \rightarrow V^*V^*$ ($V = Z$ or $W$) with three off-shell legs have been obtained in terms of Goncharov polylogarithms \citep{Birthwright:2004kk}. The results for the master integrals relevant to NLO QCD corrections to $H \rightarrow Z \gamma$ with a massive top-quark loop are also available~\citep{Bonciani:2015eua}, and the results are expressed in terms of Goncharov polylogarithms. In \citep{DiVita:2017xlr}, two-loop master integrals for QCD correction to neutral massive boson coupling to a pair of $W$ bosons were computed. 
The master integrals in these cases depend on two or three mass scales at maximum.
The full analytic results for the master integrals appearing in ${\mathcal O}(\alpha \alpha_s)$ corrections to $H W^+ W^-$ vertex are also known in terms of multiple polylogarithms \citep{Ma:2021cxg}. A few examples of one-loop triangle diagrams with arbitrary mass dependence are given in \cite{Abreu:2015zaa,Davydychev:2005nf,Tarasov:2008hw,THOOFT1979365}.
A canonical set of master integrals applicable to ${\cal O}(\alpha \alpha_s)$ corrections in $H \to Z Z^*$ decay with four mass scale dependence was studied in \cite{Wang:2019fxh}, however the analytic results for these integrals are not public.

In this article, we provide for the first time the full analytic result for all the two-loop {canonical} master integrals that contribute to ${\mathcal O}(\alpha \alpha_s)$ corrections to $H \rightarrow ZZ^*$ decay, keeping the full dependence on the mass of the top quark ($m_t$) in the loop. We also present the analytic results of all the boundary constants along with a compact set of letters called {the} alphabet. These are important for phenomenological applications as a minimal set of letters is needed for a fast numerical evaluation, and analytic boundary constants help in evaluating the master integrals with higher precision. The results also cover a more general decay $H \rightarrow Z^*Z^*$ as the number of mass scales remains the same. 
In addition, we identify the hypersurface obtained during a simultaneous rationalization of all the square roots, {occurring} because of the multiple mass scales, to be a $CY_3$ manifold. This establishes a connection between the relevant Feynman integrals for this process and integrals that occur in more symmetric quantum field theories. The choice of a set of canonical master integrals presented here is also useful for similar 2-loop massive triangles with three off-shell legs and a massive internal loop. For instance, this set of master integrals, apart from contributing to  $H \rightarrow ZZ^*$, is also useful for mixed electroweak-QCD corrections to Higgs production via the process $e^+ e^- \rightarrow H Z$ and  $e^+ e^- \rightarrow H \mu^+ \mu^-$, which will be relevant at future $e^+ e^-$ colliders \citep{Gong:2016jys,Chen:2018xau,Wang:2019fxh}. The results can also be used for precision studies in processes like $\mu^+ \mu^- \rightarrow H Z, H \ell^+\ell^-$ at future muon collider \citep{Delahaye:2019omf}.

The standard technique of evaluating multi-loop Feynman diagrams consists of first obtaining the set of master integrals using integration-by-part (IBP) identities \citep{TKACHOV198165,CHETYRKIN1981159}, then setting up a system of differential equations for the master integrals, and finally solving the differential equations using appropriate boundary conditions~\cite{KOTIKOV1991158,2003,Remiddi:1997ny,2000,doi:10.1142/S0217751X07037147,2013MWZ,2013,2015}. In general, {after bringing the system of differential {equations} into an epsilon-factorized form}, their solutions can be written in terms of iterated integrals. The iterated integrals with rational integrands can be expressed in terms of Goncharov polylogarithms \citep{10.1007/978-3-0348-9078-6_31,Goncharov:2001iea,2000HP}. However, the presence of multiple square roots in the system of differential equations poses technical challenges in the analytic computations of these master integrals \citep{Heller:2019gkq, Besier:2020hjf}. 

 The study on the effects of the presence of multiple square roots in analytic calculations of Feynman integrals has gained a lot of momentum recently \citep{1995BB,1995BBM,2013albc,2014alwsc,2018ALC, Chicherin:2020oor,dur27608,Frellesvig:2019byn,Duhr:2021fhk}. On the formal side, a lot of efforts {have} also been going on to understand the geometrical aspects of Feynman integrals. Many recent analytic computations of Feynman integrals show that the geometry is manifested by the square roots that appear in the differential equations of the master integrals or in their evaluation via direct integration~\citep{Bourjaily:2018aeq,Panzer:2014caa}. {Characterizing Feynman integrals by their geometries helps to understand the connection among various 
 theories like QCD, electroweak and super Yang-Mills (SYM) theories.} In our case, we study all the occurring square roots and comment on their rationalizability while establishing a connection with studies on geometrical aspects of Feynman integrals. { We then construct ansätze to obtain a dlog-form of the system of differential equations with a minimal set of independent one-forms involving a non-rationalizable square root.} We also present the results of all the master integrals in terms of Chen's iterated integrals~\cite{Chen:1977oja} and perform numerical checks on our results by matching them against those obtained using publicly available numerical tools.

This article is organized as follows. In section \ref{sec:notation}, we discuss the notations of the Feynman integrals and the kinematic invariants. In section \ref{subsec:canonical}, we present all the pre-canonical as well as canonical master integrals. In section \ref{sec:squareroot}, we present the details of the square root(s) and discuss their rationalization. In section \ref{sec:oneforms} we discuss all the one-forms present in the differential equations and present the alphabet. In section \ref{sec:analytic_results}, we present the analytic results of all the boundary constants for the canonical master integrals and perform checks.

\section{Two-loop Feynman diagrams and auxiliary topology}
\label{sec:notation}
\begin{figure}[!h]
\begin{center}
\includegraphics[width=12cm,height =3.3cm]{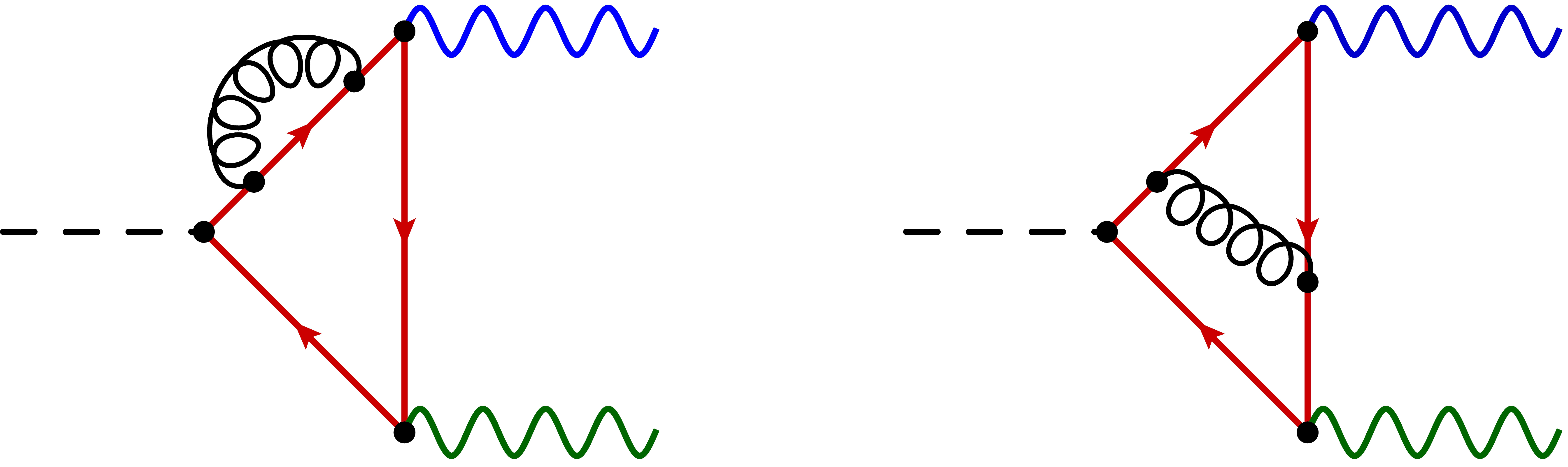}
\end{center}
\caption{Examples of Feynman diagrams contributing to the ${\mathcal O}(\alpha \alpha_s)$-corrections to the decay $H \rightarrow Z Z^*$ via a top-quark loop.
The Higgs boson is denoted by a dashed line, a top quark by a red line, $Z$ and $Z^*$ by blue and green wavy lines
and a gluon by a curly line.
}
\label{fig:Higssdecay}
\end{figure}

The representative Feynman diagrams contributing to $H(q) \rightarrow Z(p_1) + Z^*(p_2)$  at ${\mathcal O}(\alpha \alpha_s)$ are shown in figure~\ref{fig:Higssdecay}. There are two independent external momenta $p_1$ and $p_2$ corresponding to $Z$ and $Z^*$ respectively, with $p_1^2 \;=\; m_Z^2$ and $p_2^2 \;=\; m_{Z^*}^2$. The external momenta corresponding to $Z^*$ is off-shell implying 
$p_1^2 \ne p_2^2$, in general. The Higgs boson is on-shell with $q^2\; = \;(p_1+p_2)^2 \; = \; m_H^2$. 
 
\begin{figure}[!h]
\begin{center}
\includegraphics[width=9cm,height = 6cm]{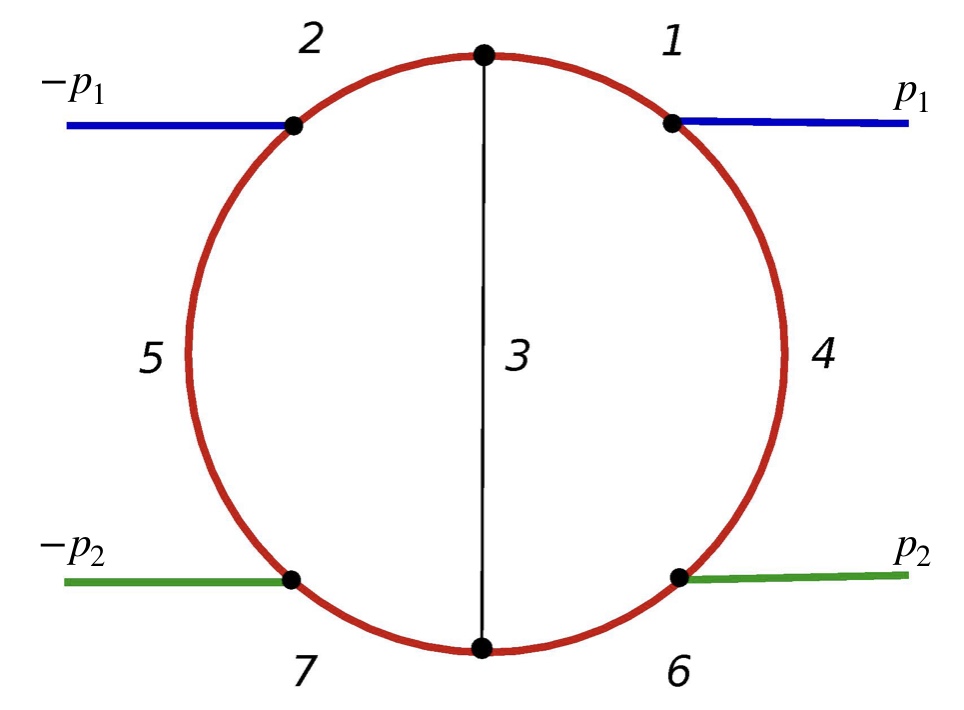}
\end{center}
	\caption{The auxiliary diagram for two-loop triangle diagrams in $H(q) \rightarrow Z(p_1) Z^*(p_2)$ decay at ${\mathcal O}(\alpha \alpha_s)$. The masses of propagators are encoded in
	the colours of the propagators; massless(light black), $m_t$(red), $m_Z$({blue}), $m_{Z^*}$({green}).
}
\label{fig:auxillary}
\end{figure}

There are  seven linearly independent scalar products involving loop momenta $k_1$ and $k_2$, and external momenta $p_1$ and $p_2$ ($k_1^2, k_2^2, k_1.k_2, k_1.p_1, k_1.p_2, k_2.p_1$ and $
k_2.p_2$) which appear in our Feynman integrals. 
The relevant two-loop Feynman diagrams, however,
have a maximum of six propagators. In order to express Feynman integrals in terms of master integrals, we  
introduce an auxiliary topology with seven propagators as shown in figure~\ref{fig:auxillary}. The corresponding integral family is given by \\
\begin{align}
\label{eqn:integral-family}
 I_{\nu_1 \nu_2 \nu_3 \nu_4 \nu_5 \nu_6 \nu_7}\left( D,p_1^2, p_2^2, m_t^2, \mu^2 \right)
 & = 
 e^{2 \gamma_E \epsilon}
 \left(\mu^2\right)^{\nu-D}
 \int \frac{d^Dk_1}{i \pi^{\frac{D}{2}}} \frac{d^Dk_2}{i \pi^{\frac{D}{2}}}
 \prod\limits_{j=1}^7 \frac{1}{ P_j^{\nu_j} }.
\end{align}
Here $\gamma_E$ denotes the Euler-Mascheroni constant, $D$ is the space-time dimension,
$\mu$ is an arbitrary scale introduced to render the Feynman integral dimensionless, 
the quantity $\nu$ is given by

\begin{align}
 \nu & = 
 \sum\limits_{j=1}^7 \nu_j.
 \end{align}
 The propagators $P_j$ are given by
 \begin{align}
 \label{eqn:propagators}
 P_1 & = -k_1^2 + m_t^2,
 &
 P_2 & = -k_2^2 + m_t^2,
 &
 P_3 & = -\left(k_1-k_2\right)^2,
 \nonumber \\
 P_4 & = -\left(k_1-p_1\right)^2 + m_t^2,
 &
 P_5 & = -\left(k_2-p_1\right)^2 + m_t^2,
 &
 P_6 & = -\left(k_1-p_1-p_2\right)^2 +m_t^2,
 \nonumber \\
 P_7 & = -\left(k_2-p_1-p_2\right)^2 + m_t^2.
 \end{align} 
We obtain the set of master integrals using the IBP program {\tt Fire} \citep{Smirnov_2008,SMIRNOV20132820,SMIRNOV2015182} combined with {\tt LiteRed} \citep{Lee:2012cn,2014LR}, which gives us 41 master integrals. The set of 41 master integrals forms a basis which 
we denote by ${\vec I}$.
For convenience, the master integrals can be classified into sectors where a sector is defined by a common set of non-zero propagators. Master integrals belonging to each sector may have different powers of the common propagators. In our case, some sectors have more than one master integral. The list of all 41 master integrals, {classified sector-wise} and following the notation in equation~\ref{eqn:integral-family}, is shown in \st{the} column 2 of table \ref{tab:master_integrals}.
We may set $\mu^2\;= \; m_t^2$, after which our master integrals depend kinematically on three dimensionless quantities defined by
\begin{align}
\label{eqn:kinematic_coordinates}
 \frac{m_{Z^*}^2}{m_t^2} = u,
 \quad
 \frac{m_H^2}{m_t^2} = v,
 \quad
 \frac{m_Z^2}{m_t^2} = w.
\end{align}

\begin{table}[!htp]
\begin{center}
\begin{tabular}{|c|l|l|c|}
\hline
 number of   & master integrals & master integrals & kinematic \\
 propagators &   basis $\vec{I}$  & basis $\vec{J}$  &dependence \\
\hline
\hline
 $2$ &  $I_{0000011}$ & $J_{1}$ & $-$ \\
\hline
\hline
 $3$ &  $I_{0000111}$ & $J_{2}$ & $u$ \\
     &  $I_{0020120}$, $I_{0010220}$ & $J_{4}, J_{5}$ & $u$ \\
     &  $I_{0100011}$ & $J_{6}$ & $v$ \\
     &  $I_{0100110}$ & $J_{7}$ & $w$ \\
     &  $I_{0120020}$, $I_{0210020}$ & $J_{12}, J_{13}$ & $v$ \\
     &  $I_{0122000}$, $I_{0212000}$ & $J_{14}, J_{15}$ & $w$ \\
\hline
\hline
 $4$ & $I_{0001111}$ & $J_{3}$ & $u$ \\
     &  $I_{0100111}$ & $J_{8}$ & $u, v, w$ \\
     &  $I_{0101011}$ & $J_{9}$ & $u, v$ \\
     & $I_{0101110}$ & $J_{10}$ & $u, w$ \\
     &  $I_{1100011}$ & $J_{16}$ & $v$ \\
     & $I_{1100110}$ & $J_{17}$ & $v, w$ \\
     &  $I_{1101100}$ & $J_{19}$ & $w$ \\
     & $I_{0110110}$, $I_{0120110}$, $I_{0110120}$ & $J_{21}, J_{22}, J_{23}$ & $u, v, w$ \\
     & $I_{0111001}$, $I_{0121001}$, $I_{0112001}$ & $J_{24}, J_{25}, J_{26}$ & $u, v, w$ \\
      & $I_{0111010}$, $I_{0121010}$, $I_{0211010}$ & $J_{27}, J_{28}, J_{29}$ & $u, v, w$ \\
\hline
\hline
 $5$ &  $I_{0101111}$ & $J_{11}$ & $u, v, w$ \\
     &  $I_{1100111}$ & $J_{18}$ & $u, v, w$ \\
     &  $I_{1101101}$ & $J_{20}$ & $u, v, w$ \\
     &  $I_{0111011}, I_{0211011}, I_{0112011},$ & $J_{30}, J_{31}, J_{32},$ & $u, v, w$ \\
     &       $I_{0212011}$ & $J_{33}$ & \\
     &  $I_{0111110}, I_{0111120}, I_{0211110},$ & $J_{34}, J_{35}, J_{36},$ & $u, v, w$ \\
     &       $I_{0211120}$ & $J_{37}$ & \\
     & $I_{1110110}, I_{1110210}, I_{1110120},$ & $J_{38}, J_{39}, J_{40},$ & $u, v, w$ \\
     &       $I_{1110220}$ & $J_{41}$ & \\
\hline
\end{tabular}
\end{center}
\caption{Overview of the set of master integrals for two-loop triangle diagrams in $H \to Z Z^*$ decay.
The first column gives the number of non-zero propagators which appear in the master integrals, the 
second column lists the master integrals in the basis $\vec{I}$,
the third column lists corresponding master integrals in the basis $\vec{J}$.
The last column gives the kinematic dependence of master integrals in 
terms of dimensionless quantities defined in equation~\ref{eqn:kinematic_coordinates}. 
}
\label{tab:master_integrals}
\end{table}

\section{An epsilon-form for the master integrals}
\label{subsec:canonical}

We set up a linear system of differential equations of the form $d\vec I =  A \vec I$ for all the master integrals of basis $\vec I$  by taking { their derivatives with respect to the kinematic invariants ($u$, $v$ and $w$) and using IBP identities.} 
The $41\times 41$ matrix $A$ depends on kinematic invariants and space-time dimensions ($D$).
The system of differential equations for the master integrals is solved iteratively at each order in the dimensional regularization parameter $\epsilon$ after { switching from basis $\vec I$ to a canonical basis $\vec J$} such that the system of differential equations has an epsilon-form \citep{2013} given by 
\begin{equation}
    d\vec J = \epsilon \tilde{A} \vec J.
    \label{eqn:epsilon-form}
\end{equation}
Here matrix $\tilde{A}$ is independent of the dimensional regulator $\epsilon$. We aim to find a transformation matrix that brings 
the pre-canonical basis ${\vec I}$ to a canonical basis ${\vec J}$ leading to a differential system given in equation~\ref{eqn:epsilon-form}.

\subsection{Finding the canonical basis}
Arranging the master integrals sector-wise gives a block-diagonal form for their differential equations. We also use a ``bottom-up'' approach where we bring the differential equations of a lower sector to a canonical form using a coordinate system suitable to that sector before moving on to a higher sector. For some of the master integrals, it is convenient to work in $D$ = $2- 2\epsilon$ dimensions. We use the dimension shifting operator $\textbf{D}^-$ to shift the integrals from $4- 2\epsilon$ to $2- 2\epsilon$. We express such integrals as a linear combination of master integrals in $D= 4- 2\epsilon$ dimensions using the dimensional shift relations~\citep{Lee:2012cn}. To find a transformation matrix that brings our pre-canonical basis to a canonical form, we use the information of the maximal cuts of each of the sector \citep{Harley:2017qut,Weinzierl:2022eaz,Primo:2016ebd,Frellesvig:2017aai,Bosma:2017ens}. Following the same method, as described in ~\citep{2018ALC,Chaubey:2019lum,Frellesvig:2019byn}, we first construct an ansatz for a canonical basis using the information from the maximal cut of the sector in consideration. This fixes the diagonal part of the differential equations for that sector. We then include the contribution from the sub-sectors to complete the construction of a canonical basis for that sector. Finally, we repeat this process to reach the top sector and construct a canonical basis for the full system. Our canonical basis $\vec{J}$, in terms of dimensionless quantities $u, v \text{ and } w$, is given by the following transformations.

\begin{align*}
J_{1}\;&= \; \epsilon^2 \;\; \textbf{D}^{-} I_{0000011},\\
J_{2}\;&= \; \epsilon^2\; \sqrt{-u\;\left(4-u\right)} \; \textbf{D}^{-} I_{0000111},\\
J_{3}\;&= \; \epsilon^2\;u\;\left(4-u\right)\; \textbf{D}^{-} I_{0001111},\\
J_{4}\;&= \; \epsilon^2 \sqrt{-u\;\left(4-u\right)}  \left[ I_{0020120}+\frac{1}{2} I_{0010220}\right],\\
J_{5}\;&= \; \epsilon^2\; u \; I_{0010220},\\
J_{6}\;&= \; \epsilon^2 \; \sqrt{-v\;\left(4-v\right)} \; \textbf{D}^{-} I_{0 1 0 0 0 1 1},\\
J_{7}\;&= \; \epsilon^2\;\sqrt{-w\;\left(4-w\right)} \;\textbf{D}^{-} I_{0 1 0 0 1 1 0},\\
J_{8}\;&= \;  \frac{\epsilon^2}{2\;\sqrt{\lambda}}\bigg[ \; 2\left(\lambda+ u\; v\; w\right)\;\textbf{D}^{-}I_{0 1 0 0 1 1 1} + w\;(u + v - w) \;\textbf{D}^{-}I_{0 1 0 0 1 1 0} \\&+ v (u - v + w) \;\textbf{D}^{-}I_{0 1 0 0 0 1 1} + u\;(-u + v + w)\; \textbf{D}^{-} I_{0 0 0 0 1 1 1} \bigg],\\
J_{9}\;&= \; \epsilon^2\;  \sqrt{u\;\left(4-u\right)}\; \sqrt{v\;\left(4-v\right)}\;\textbf{D}^{-}I_{0 1 0 1 0 1 1} ,\\
J_{10}\;&= \; \epsilon^2\; \sqrt{u\;\left(4-u\right)}\;\sqrt{w\;\left(4-w\right)} \;\textbf{D}^{-}I_{0 1 0 1 1 1 0},\\
J_{11}\;&= \; \frac{\epsilon^2 }{2}\sqrt{\frac{-u\;\left(4-u\right)}{\lambda}}\;\bigg[ 2\left(\lambda+ u\; v\; w\right) \;\textbf{D}^{-}I_{0 1 0 1 1 1 1}+  w\;(u + v - w)\;\textbf{D}^{-}I_{0 1 0 1 1 1 0} \\&+ v\;(u - v + w) \;\textbf{D}^{-}I_{0 1 0 1 0 1 1} +  u\;(-u + v + w)\; \textbf{D}^{-} I_{0001111} \bigg],\\
J_{12}\;&= \; \epsilon^2\; \sqrt{-v\;\left(4-v\right)}  \left[I_{0 1 2 0 0 2 0}+\frac{1}{2}I_{0 2 1 0 0 2 0}\right],\\
J_{13}\;&= \; \epsilon^2 \; v\; I_{0 2 1 0 0 2 0},\\
J_{14}\;&= \; \epsilon^2 \; \sqrt{-w\;\left(4-w\right)} \left[I_{0 1 2 2 0 0 0}+\frac{1}{2}I_{0 2 1 2 0 0 0}\right],\\
J_{15}\;&= \; \epsilon^2 \; w\; I_{0 2 1 2 0 0 0},\\
J_{16}\;&= \; \epsilon^2 \; v\;\left(4-v\right)\;\textbf{D}^{-} I_{1 1 0 0 0 1 1},\\
J_{17}\;&= \; \epsilon^2 \;  \sqrt{v\;\left(4-v\right)}\;\sqrt{w\;\left(4-w\right)}\;\textbf{D}^{-} I_{1 1 0 0 1 1 0},\\
J_{18}\;&= \;  \frac{\epsilon^2}{2}\sqrt{\frac{-v\;\left(4-v\right)}{\lambda}}\;\bigg[2\left(\lambda+ u\; v\; w\right) \;\textbf{D}^{-} I_{1 1 0 0 1 1 1}+  w\;(u + v - w) \;\textbf{D}^{-} I_{1 1 0 0 1 1 0} \\&+ v\;(u - v + w)\;\textbf{D}^{-} I_{1 1 0 0 0 1 1}+  u\;(-u + v + w)\;\textbf{D}^{-}I_{0 1 0 1 0 1 1} \bigg],\\
J_{19}\;&= \; \epsilon^2\;w\;\left(4-w\right)\;\textbf{D}^{-} I_{1 1 0 1 1 0 0},\\
J_{20}\;&= \; \frac{\epsilon^2}{2}\sqrt{\frac{-w\;\left(4-w\right)}{\lambda}}\;\bigg[ 2\left(\lambda+ u\; v\; w\right) \;\textbf{D}^{-} I_{1 1 0 1 1 0 1} +  w\;(u + v - w) \;\textbf{D}^{-} I_{1 1 0 1 1 0 0} \\ & +v\;(u - v + w) \;\textbf{D}^{-} I_{1 1 0 0 1 1 0} + u\;(-u + v + w) \;\textbf{D}^{-}I_{0 1 0 1 1 1 0}\bigg] ,\\
J_{21}\;&= \;\epsilon^3\; \sqrt{\lambda}\; I_{0 1 2 0 1 1 0},\\
J_{22}\;&= \; \epsilon^3\; \sqrt{\lambda}\;I_{0 1 1 0 1 2 0},\\
J_{23}\;&= \; \epsilon^2\;\frac{\sqrt{-w\;(4-w)}}{w\;(u + v - w)}\;\bigg[\left(\lambda+ u\; v\; w\right)\;\textbf{D}^{-} I_{0 1 1 0 1 1 0} - 2\; \epsilon \; \lambda \;( I_{0 1 2 0 1 1 0} + \frac{1}{2} I_{0 1 1 0 1 2 0} )\\&+  v\;(u - v + w)\;( I_{0 1 2 0 0 2 0} + \frac{1}{2} I_{0 2 1 0 0 2 0}) + u\;(-u + v + w)\; (I_{0 0 2 0 1 2 0}+\frac{1}{2} I_{0 0 1 0 2 2 0})\bigg],\\
J_{24}\;&= \; \epsilon^3\;\sqrt{\lambda}\;I_{0 1 2 1 0 0 1},\\
J_{25}\;&= \; \epsilon^3\; \sqrt{\lambda}\; I_{0 1 1 2 0 0 1},\\
J_{26}\;&= \; \; \epsilon^2\;\frac{\sqrt{-v\;(4-v)}}{v\;(-u + v - w)}\;\bigg[\left(\lambda+ u\; v\; w\right)\;\textbf{D}^{-} I_{0 1 1 1 0 0 1} - 2\; \epsilon \; \lambda \;( I_{0 1 2 1 0 0 1} + \frac{1}{2} I_{0 1 1 2 0 0 1} )\\&+  w\;(u + v - w)\;( I_{0 1 2 2 0 0 0} + \frac{1}{2} I_{0 2 1 2 0 0 0}) + u\;(-u + v + w)\; (I_{0 0 2 0 1 2 0}+\frac{1}{2} I_{0 0 1 0 2 2 0})\bigg],\\
J_{27}\;&= \; \epsilon^3 \; \sqrt{\lambda}\;I_{0 1 2 1 0 1 0},\\
J_{28}\;&= \; \epsilon^3\; \sqrt{\lambda}\;I_{0 2 1 1 0 1 0},\\
J_{29}\;& =\; \; \epsilon^2\;\frac{\sqrt{-u\;(4-u)}}{u\;(-u + v + w)}\;\bigg[\left(\lambda+ u\; v\; w\right)\;\textbf{D}^{-} I_{0 1 1 1 0 1 0} - 2\;\epsilon \; \lambda \;( I_{0 1 2 1 0 1 0} + \frac{1}{2} I_{0 2 1 1 0 1 0} )\\&+  w\;(u + v - w)\;( I_{0 1 2 2 0 0 0} + \frac{1}{2} I_{0 2 1 2 0 0 0}) + v\;(u - v + w)\; (I_{0 1 2 0 0 2 0}+\frac{1}{2} I_{0 2 1 0 0 2 0})\bigg],\\
J_{30}\;& =\; \epsilon^4 \; \sqrt{\lambda}\; I_{0 1 1 1 0 1 1},\\
J_{31}\;&= \; \epsilon^3\;\sqrt{\lambda}\; \sqrt{-v\;\left(4-v\right)}\; I_{0 2 1 1 0 1 1},\\
J_{32}\;&= \; \epsilon^3\;\sqrt{\lambda}\; \sqrt{-u\;\left(4-u\right)}\; I_{0 1 1 2 0 1 1},\\
J_{33}\;&= \; \epsilon^2\;\bigg[\left(\lambda+ u\; v\; w\right) \; I_{0 2 1 2 0 1 1} + \epsilon\; u\;(-u + v + w) \; I_{0 1 1 2 0 1 1}+ \epsilon \;v\;(u - v + w) \; I_{0 2 1 1 0 1 1}\\&+(w + \frac{1}{2}\; u \;(-2 + v) - v)\;\textbf{D}^{-} I_{0 1 0 1 0 1 1}\bigg],\\
J_{34}\;&= \; \epsilon^4\; \sqrt{\lambda}\;I_{0 1 1 1 1 1 0},\\
J_{35}\;&= \; \epsilon^3\;\sqrt{\lambda}\;\sqrt{-u\;\left(4-u\right)}\;I_{0 1 1 1 1 2 0},\\
J_{36}\;&= \; \epsilon^3 \sqrt{\lambda}\;\sqrt{-w\;\left(4-w\right)}\;I_{0 2 1 1 1 1 0},\\
J_{37}\;&= \; \epsilon^2 \;\bigg[\left(\lambda+ u\; v\; w\right) \; I_{0 2 1 1 1 2 0} +\epsilon\; w\;(u + v - w) \; I_{0 2 1 1 1 1 0}+ \epsilon\;u\;(-u + v + w) \; I_{0 1 1 1 1 2 0}\\&+(v + \frac{1}{2}\; u (-2 + w) - w)\;\textbf{D}^{-} I_{0 1 0 1 1 1 0}\bigg],\\
J_{38}\;&= \; \epsilon^4\; \sqrt{\lambda}\; I_{1 1 1 0 1 1 0},\\
J_{39}\;&= \; \epsilon^3 \;\sqrt{\lambda}\;\sqrt{-w\;\left(4-w\right)}\; I_{1 1 1 0 2 1 0},\\
J_{40}\;&= \; \epsilon^3\; \sqrt{\lambda}\;\sqrt{-v\;\left(4-v\right)}\;I_{1 1 1 0 1 2 0},\\ 
J_{41}\;&= \; \epsilon^2\;\bigg[ \left(\lambda+ u\; v\; w\right) \; I_{1 1 1 0 2 2 0} +\epsilon\; v\;(u - v + w) \; I_{1 1 1 0 1 2 0}+ \epsilon\;w\;(u + v - w) \;I_{1 1 1 0 2 1 0}\\&+(u + \frac{1}{2}\; v \;(-2 + w) - w)\;\textbf{D}^{-} I_{1 1 0 0 1 1 0}\bigg].
\end{align*}
In the above, $\lambda(u,v,w)$ is the K\"allen function defined by
\begin{equation}
 \lambda(u,v,w)  =  u^2 + v^2 + w^2 - 2 u v - 2 v w - 2 w u.
\end{equation}

Note that in terms of the coordinates of equation~\ref{eqn:kinematic_coordinates}, the system of differential 
equations contains multiple square roots. In particular, the following square roots appear 
\begin{equation}
 \sqrt{-u\left(-4+u\right)},
 \quad 
 \sqrt{-v\left(-4+v\right)},
 \quad
 \sqrt{-w\left(-4+w\right)}
  \quad \mbox{and} \quad 
 \sqrt{\lambda\left(u,v,w\right)}.
 \label{eqn:sqrt}
\end{equation}
A non-trivial task is to find a coordinate system which rationalizes all these square roots simultaneously. We introduce the following set of transformations that takes us from $(u, v, w)$ to $(x, y, z)$ and rationalizes the first three square roots in equation~\ref{eqn:sqrt}
\begin{equation}
    \frac{m_{Z^*}^2}{m_t^2} \; = \; u \; = \; - \frac{\left(1-x\right)^2}{x}
    ,\quad
     \frac{m_{H}^2}{m_t^2} \; = \; v \; = \; - \frac{\left(1-y\right)^2}{y}
     ,\quad
     \frac{m_{Z}^2}{m_t^2} \; = \; w \; = \; - \frac{\left(1-z\right)^2}{z}.
     \label{eqn:xyz}
\end{equation}
These transformations are ubiquitous in literature \citep{1999fkv}. After using these set of transformation we are left with only one square root $\sqrt{\lambda\left(u,v,w\right)}$, which in the new coordinate system becomes
\begin{equation}
\small
 r =  \sqrt{\begin{aligned}&(y^2 z^2 + x^4 y^2 z^2 - 2 x y z (z + y (1 + z (-2 + y + z)))- 2 x^3 y z (z + y (1 + z (-2 + y + z)))\\& + x^2 (-2 y (-1 + z)^2 z - 2 y^3 (-1 + z)^2 z + z^2 + y^4 z^2 + y^2 (1 + z (4 + z (-6 + z (4 + z)))))).\end{aligned}}
 \label{eqn:the_sqrt}
\end{equation}
This square root $r$ cannot be rationalized by doing further coordinate transformations. We discuss the appearance of square roots {and their rationalizability} in more detail in the next section. 

\begin{figure}[ht!]
\centering
\hspace{0.8cm}\subfloat[$J_1$]{\includegraphics[width = 1.5in]{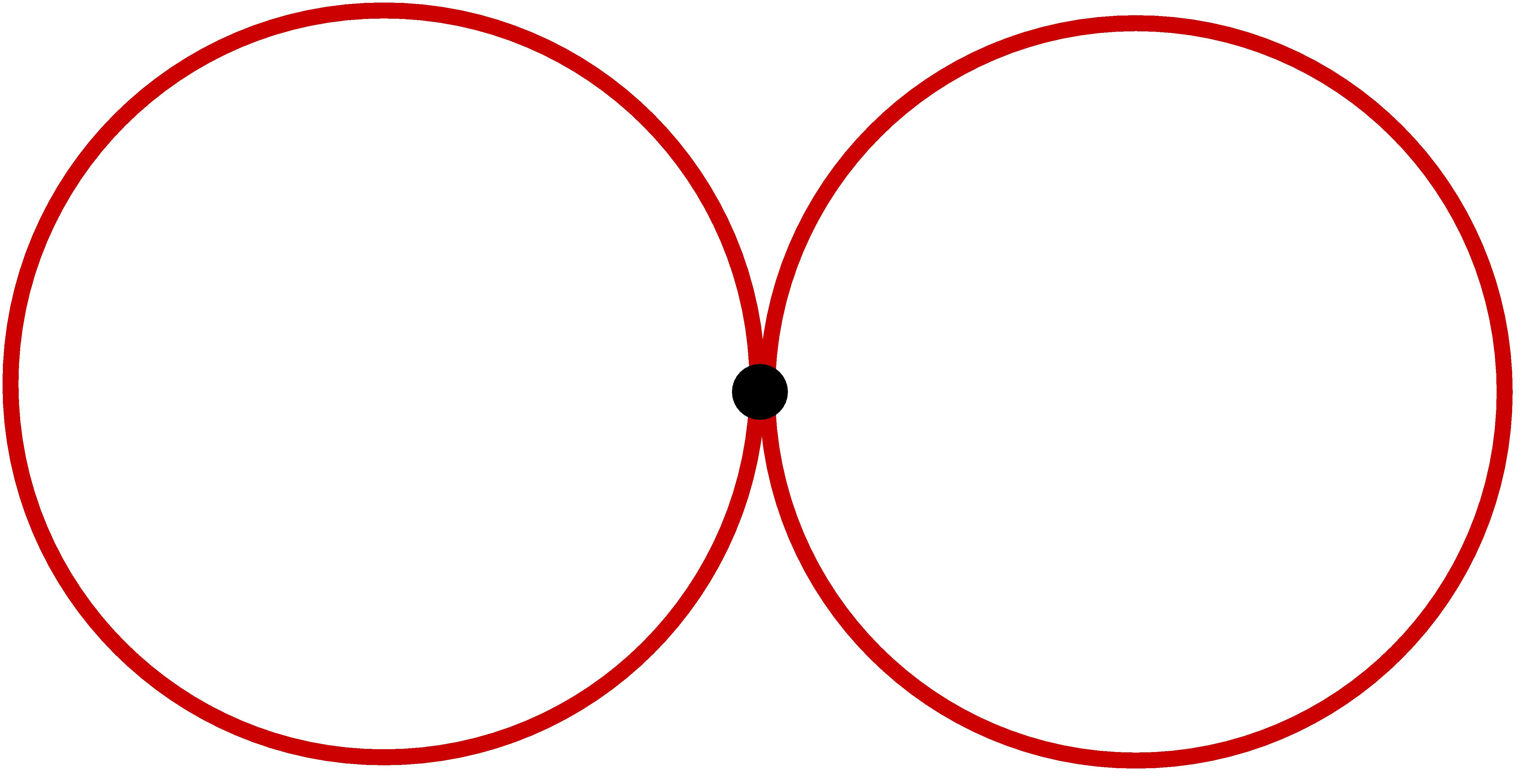}} \hspace{1cm}
\subfloat[$J_2, J_6, J_7$]{\includegraphics[width = 2in]{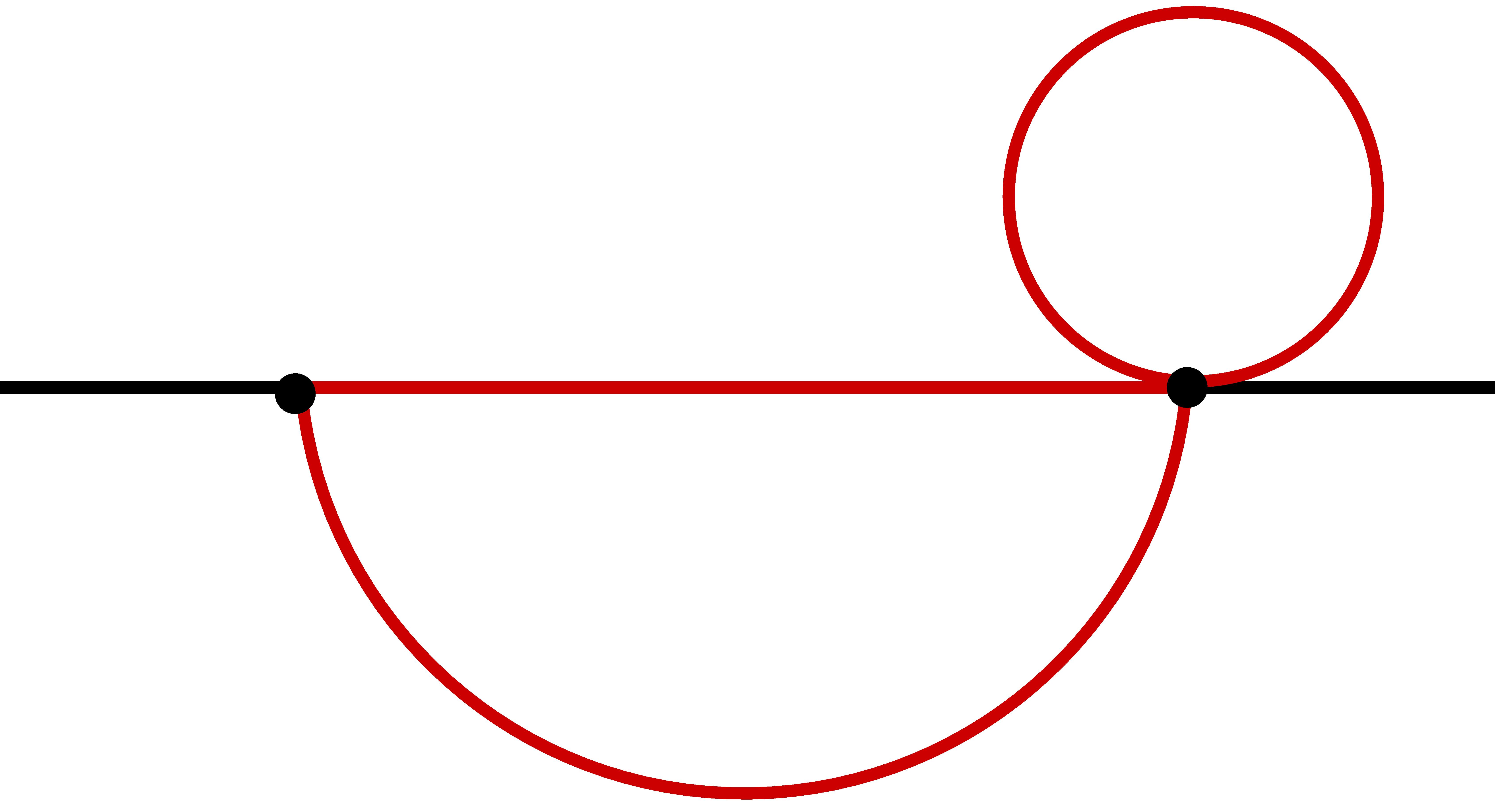}}\\
\subfloat[$J_3, J_{16}, J_{19}$]{\includegraphics[width = 2in]{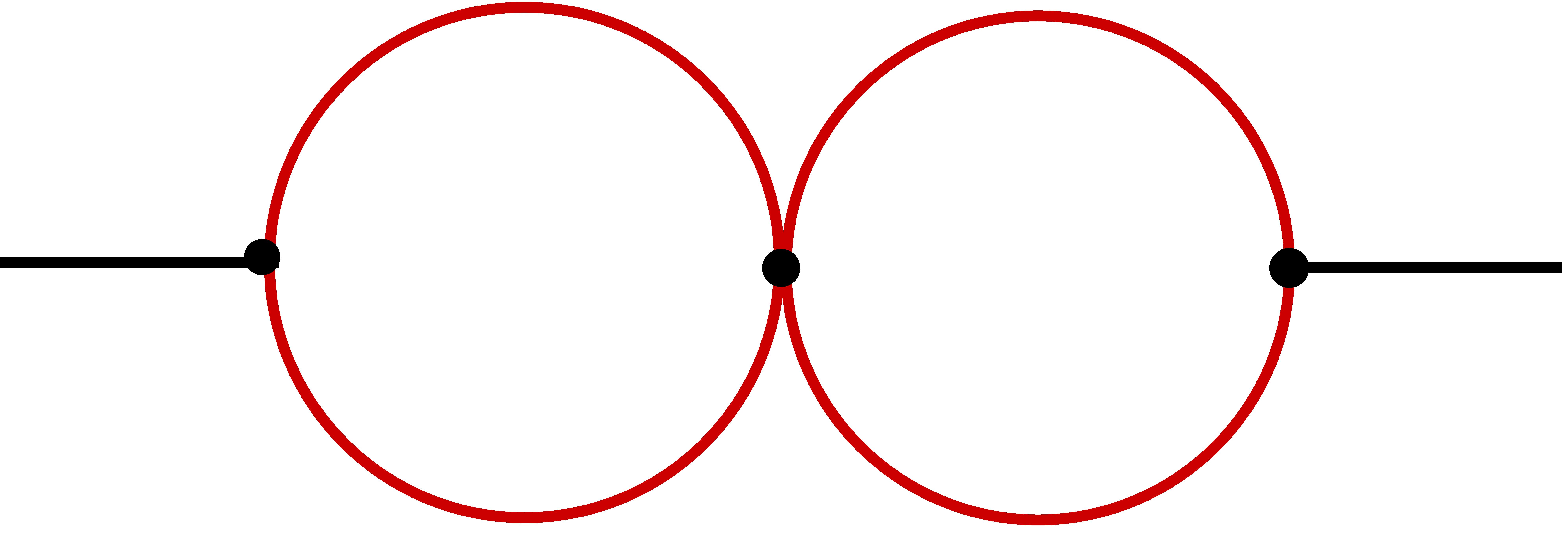}}
\hspace{1cm}
\subfloat[$J_4, J_{5}, J_{12}, J_{13}, J_{14}, J_{15}$]{\includegraphics[width = 1.5in]{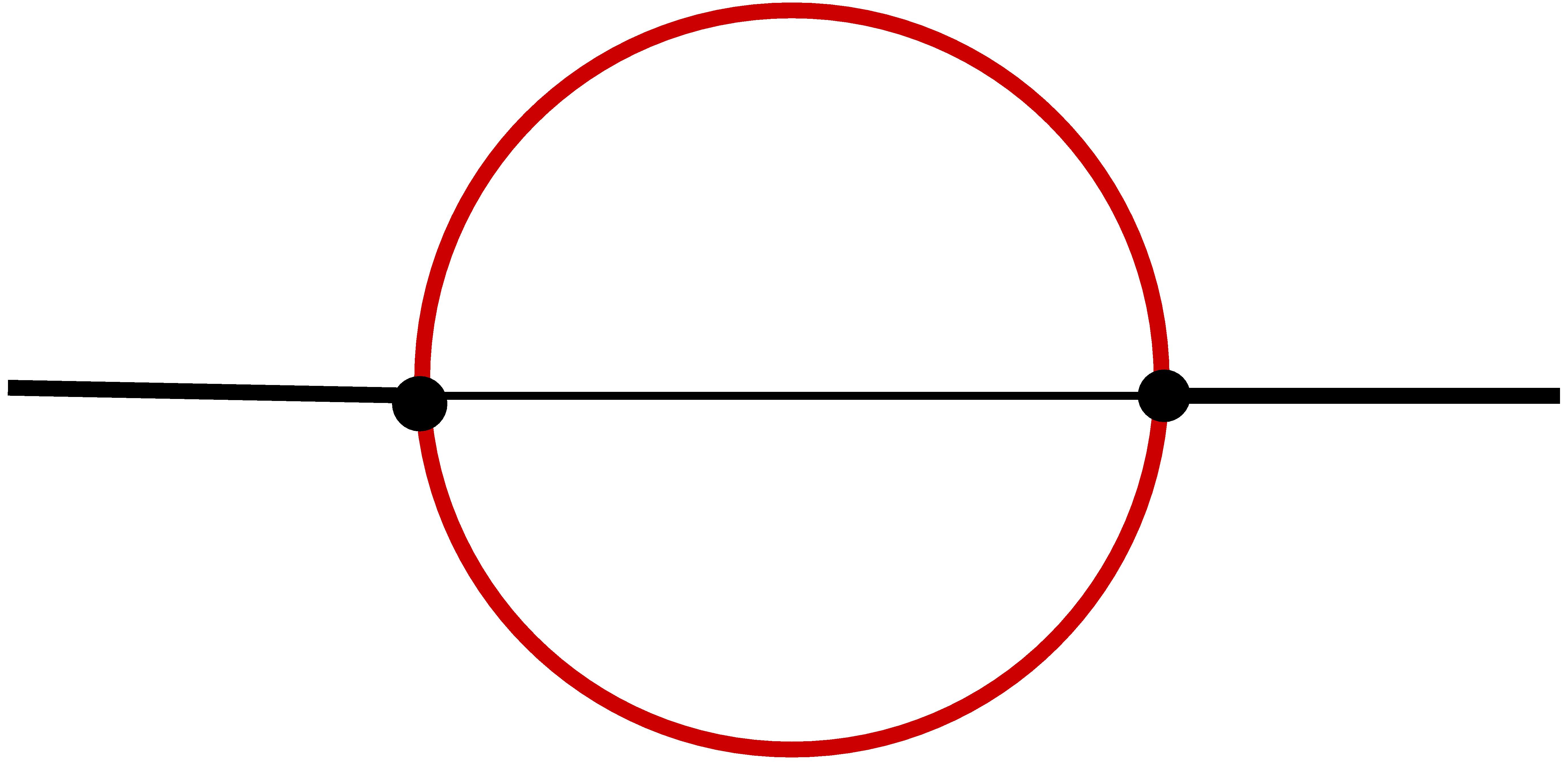}} \\
\hspace{0.5cm}\subfloat[$J_8 $]{\includegraphics[width = 1.5in]{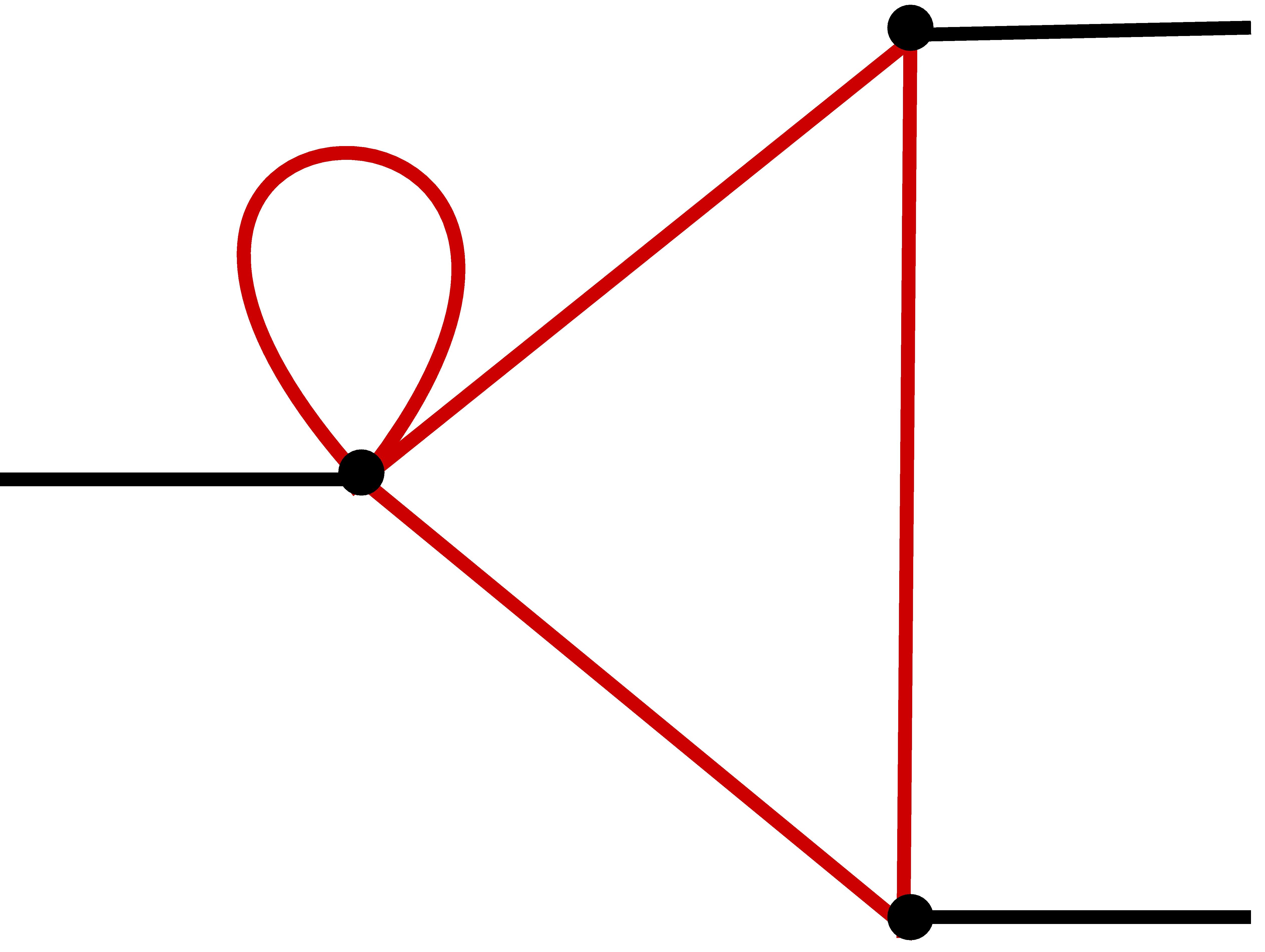}} 
\hspace{2cm}
\subfloat[$J_9, J_{10}, J_{17} $]{\includegraphics[width = 2in]{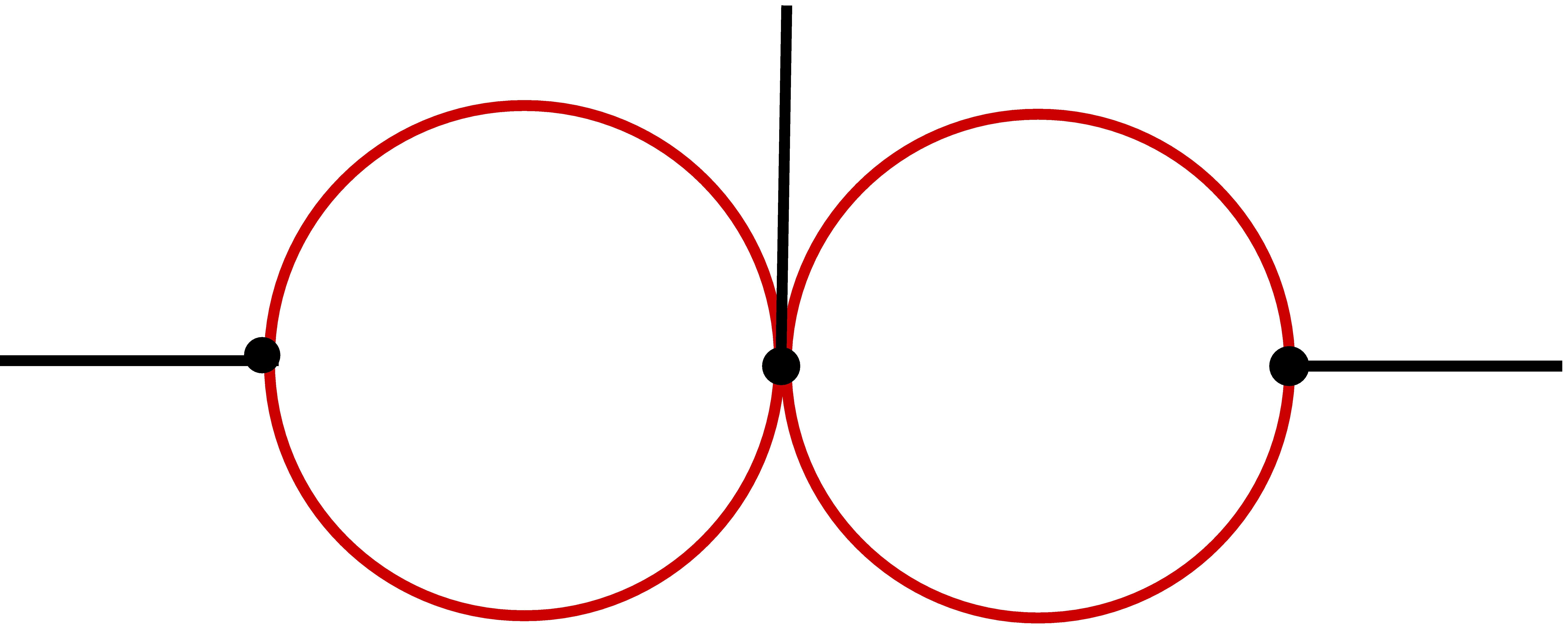}}\\
\subfloat[$J_{11}, J_{18}, J_{20}$]{\includegraphics[width = 2in]{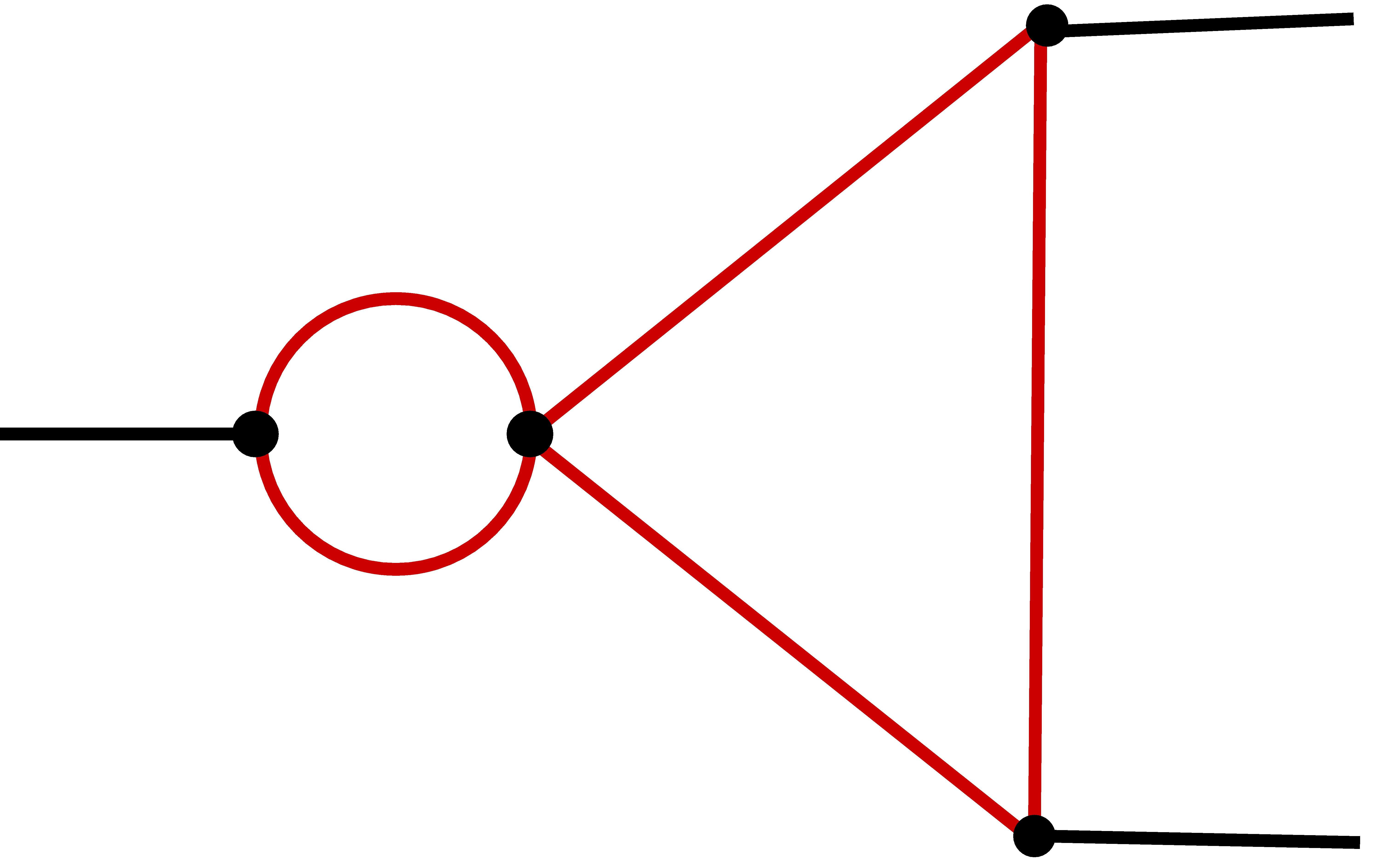}}
\hspace{2cm}
\subfloat[$J_{21}-J_{29}$]{\includegraphics[width = 1.5in]{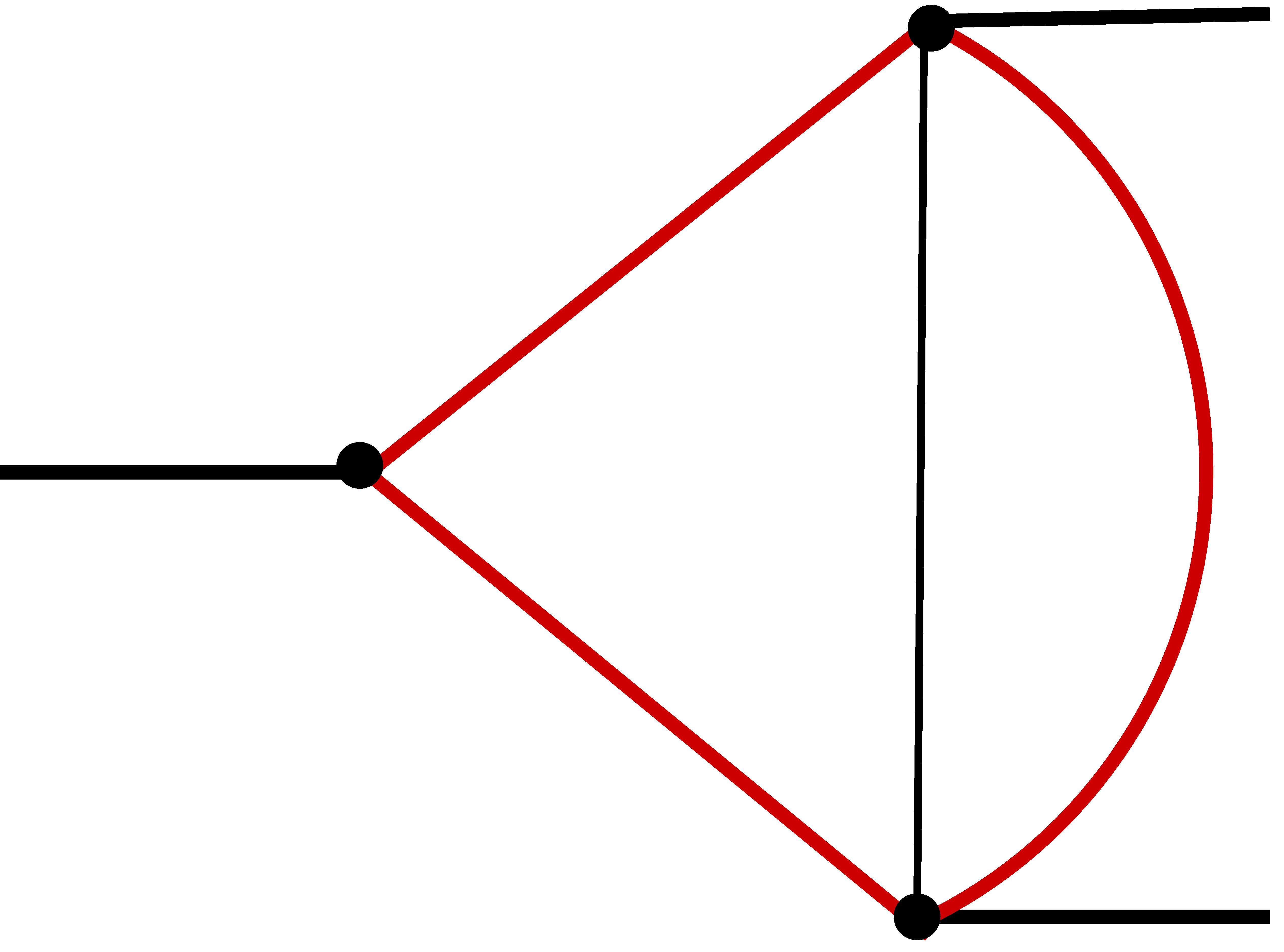}} \\
\subfloat[$J_{30}-J_{41}$]{\includegraphics[width = 2.2in]{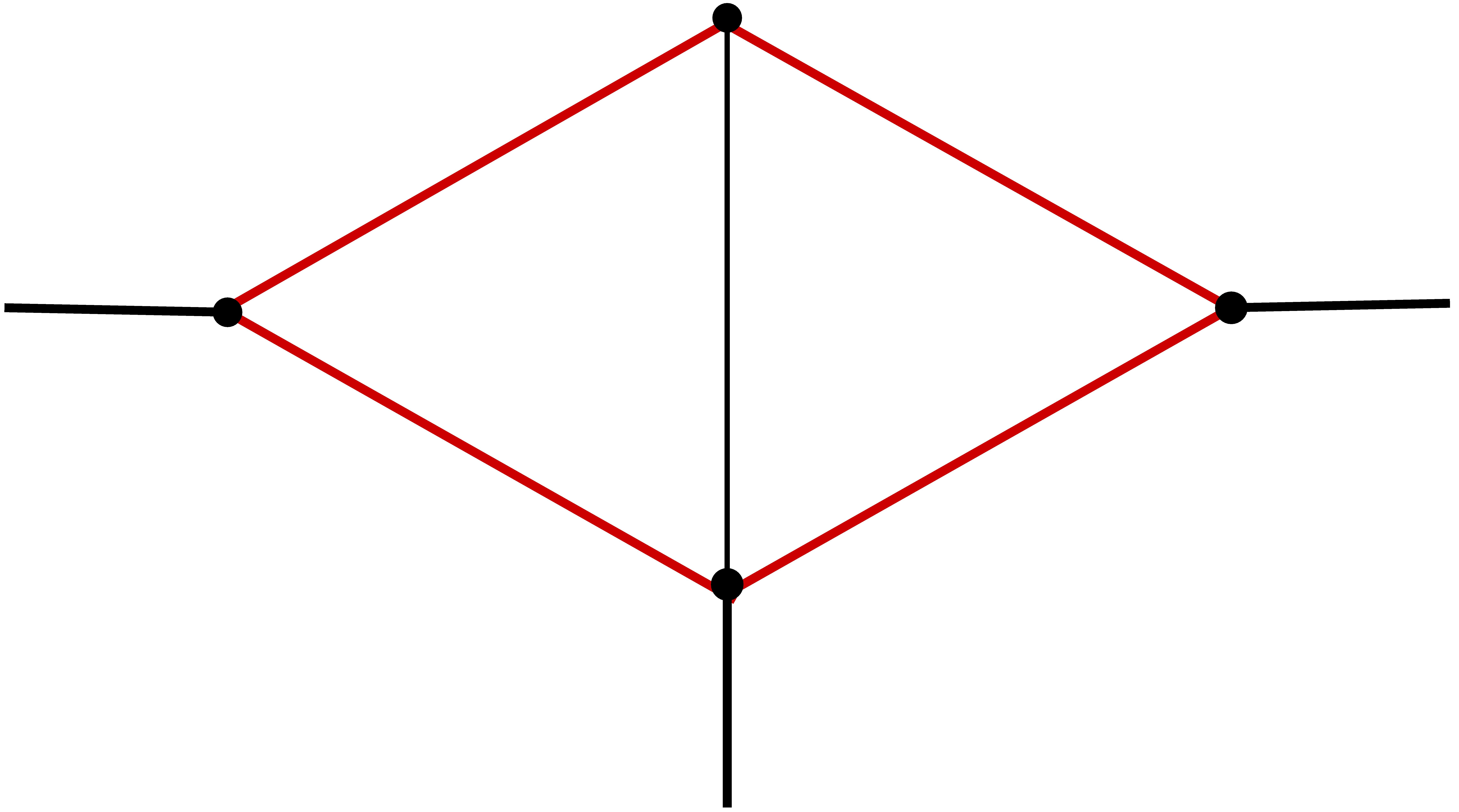}} 
\caption{Topologies relevant to two-loop {three-point} diagrams in $H\rightarrow Z Z^*$ decay at ${\mathcal O}(\alpha \alpha_s)$. Light black lines represent massless propagators. The caption of each topology denotes the set of canonical master integrals belonging to that topology.}
\label{fig:topologies}
\end{figure}

Sector-wise classification of the master integrals of canonical basis and their kinematic dependence {is given in {columns} 3 and 4, respectively,} of table~\ref{tab:master_integrals}. The master integrals from higher sectors have more propagators and are usually more complicated to solve than the master 
integrals from the lower sectors. We also look at the topology of each sector, where multiple sectors may correspond to the same topology. The form of canonical basis can be recycled within sectors belonging to the same topology since these sectors are usually obtained by a permutation of the external legs. In our case, we have 9 
master topologies which are shown in figure \ref{fig:topologies}. Apart from being a guiding principle in the construction of a canonical basis, the classification of integrals in terms of master topologies also helps 
in fixing the boundary constants {while} solving the differential equations. The boundary constants are discussed in section~\ref{sec:analytic_results}. 

\subsection{The square root}
\label{sec:squareroot}
Due to the presence of many mass scales in our system, the differential equations contain { 4 square roots in the coordinate system given in equation \ref{eqn:kinematic_coordinates}. However, our choice of variables given in equation \ref{eqn:xyz} rationalizes 3 out of 4 square roots present in the system.} The fourth square root is $r = \sqrt{P(x,y,z)}$ given in equation \ref{eqn:the_sqrt}. This square root is present in the analytic expressions of the master integrals $J_{8}$, $J_{11}$, $J_{18}$, $J_{20}-J_{41}$. In general, one can try to perform a change of variables using 
\begin{equation}
 x = \psi_1(x_1),\; \; y=\psi_2(x_2),\; \; z = \psi_3(x_3),  
\end{equation}
with rational functions $\psi_1$, $\psi_2$ and $\psi_3$, such that $P(\psi_1(x_1),\psi_2(x_2),\psi_3(x_3))$ becomes a perfect square. Several algorithms have been devised in the past to identify rational parametrizations for square roots that appear in certain Feynman integral calculations~\cite{Besier:2018jen}, or to demonstrate that rationalization is not achievable \citep{dino2018bhabha1,2020BMWP,Besier:2020klg,Besier:2020hjf,Festi:2021tyq}. In particular, it is simpler to find a transformation for rationalizing the square roots with one-variable dependence. The condition to find such a parametrization is that the degree of the polynomial involved must be less than or equal to 2. However, in some cases, the presence of a single square root of a multi-variable polynomial or a polynomial with a degree greater than 2 makes the subject of rationalizability more complicated \citep{Besier:2018jen,2020BMWP,Besier:2020klg, Besier:2020hjf, Festi:2021tyq}. There exists a number of examples where the analytic computation of Feynman integrals involve non-rationalizable square roots of degree $\geq$ 3, for which the  results are expressed in terms of a more general class of functions such as elliptic multi-polylogarithms (eMPLs) (iterated integrals over moduli space of torus) \citep{2005lr,Muller-Stach:2012tgz,brown2013multiple13,2015bsvp,2013albc,2014alwsc,Adams:2015gva,2016albw,2016albcs,Adams:2017ejb,2017alcew,2017bcsa,2018bjdcd,2018bjdcdf,Adams:2018yfj,2018bjdt,2018sw,2019bt,2018ALC,2018hi,2020bcm,2019fp,2019rp,weinzierl2019simple,Duhr:2021fhk} and $K3$ surfaces  \citep{Besier:2018jen,dino2018bhabha1,2012bfs,Bloch:2016izu,Besier:2019hqd}. Recent studies show that even more complicated functions are expected to appear for massive cases at higher orders in scattering amplitudes in various QFTs which can be expressed as integrals over manifolds in higher dimensions. In many cases, these manifolds are found to be Calabi-Yau \cite{brown2010periods,2012bfs,2019bjl,2018bjlh,dino2018bhabha1,Besier:2019hqd,2020bjlm}. In order to study the square root $r$ and to understand its connection with a Calabi-Yau threefold variety, discussed before in the literature~\citep{2020bjlm,2019bjl}, we proceed as follows. We start by noting that $P(x,y,z)$ in the square root $r$ we obtain after simultaneously rationalizing three out of the four square roots of equation~\ref{eqn:sqrt}

\begin{equation}
   \begin{aligned}  P(x,y,z) &= r^2 \\&= y^2 z^2 + x^4 y^2 z^2 - 2 x y z (z + y (1 + z (-2 + y + z)))- 2 x^3 y z (z + y (1 + z (-2 + y + z)))\\& + x^2 (-2 y (-1 + z)^2 z - 2 y^3 (-1 + z)^2 z + z^2 + y^4 z^2 + y^2 (1 + z (4 + z (-6 + z (4 + z))))) \end{aligned}
\end{equation}
is a degree 8 in-homogeneous polynomial with no repeated roots. Then we assign weight one to all the three coordinates $x$, $y$ and $z$ to realize this surface as a projective variety. In the next step, we introduce an additional auxiliary coordinate `$l$' to homogenize $P(x,y,z)$. As a result, we obtain a homogeneous polynomial $P_{8}(x,y,z,l)$ of overall degree 8 in 4 variables. Writing this hypersurface as
\begin{equation}
      Q(x,y,z,l,r) = r^2 - P^2_{8}(x,y,z,l) = 0,
\end{equation}
we can identify it as a degree 8 hypersurface in 4-dimensional weighted projective space $\mathbb{WP}^{1,1,1,1,4}$, with weight 4 assigned to $r$. The sum of the 
weights of this weighted projective space is 8 which is exactly equal to the degree of the polynomial $Q$, known as a Calabi-Yau threefold $CY_3$ in $\mathbb{WP}^{1,1,1,1,4}$~\citep{2020bjlm}\footnote{We thank Matthias Wilhelm \& Matthew von Hippel for discussions regarding this.}. With this identification, one can obtain the Hodge structure and Euler characteristic that characterize the Calabi-Yau manifold. {After establishing this interesting connection, we proceed to bring the epsilon-factorized differential equation system into a dlog-form, and express our results for the master integrals in terms of Chen's iterated integrals with dlog one-form kernels. 
Even though from a mathematical point of view, it might be possible to obtain a representation {of these iterated integrals} in terms of elliptic multiple polylogarithms (eMPLs)~\citep{Duhr:2021fhk} with such kernels, we would like to keep this discussion for the future and proceed keeping the phenomenological 
applications of our results in mind.
More details on one-forms and {the}
motivation behind using iterated integrals over dlog one-forms for phenomenological 
applications {are} provided in the following section.}

\subsection{dlog one-forms}
\label{sec:oneforms}

We Taylor expand the canonical basis integrals $J_k$ around $\epsilon=0$ as 
\begin{equation}
 J_k
 =
 \sum\limits_{j=0}^\infty \epsilon^j J_k^{(j)}.\label{eqn:expandedJ}
\end{equation}
Putting it in the $\epsilon$-factorized differential equation of \ref{eqn:epsilon-form}, allows us to express each $J_k^{(j)}$ 
in terms of Chen's iterated integrals up to a boundary constant.
Chen's iterated integrals are defined as follows.
Let there be a path $\gamma$ on an $n$-dimensional manifold $M$, $\gamma  : [0,1]\rightarrow M $, where $\gamma(0)$ is the starting point and $\gamma(1)$ is the endpoint. Let the set of differential one-forms be denoted by ${\omega_i}$ and their pullback to the interval $[0,1]$ be given as $\gamma^ * \omega_j=f_j(\lambda) d\lambda$. The $k$-fold iterated integral over the one-forms is defined as
\begin{align}
I_\gamma (\omega_1,..., \omega_k; \lambda)& =\int_0^\lambda d\lambda_1 f_1(\lambda_1) \int_0^{\lambda_1} d \lambda_2 f_2 (\lambda_2) ... \int_0^{\lambda_{k-1}} d \lambda_k f_k (\lambda_k)\nonumber\\
&= \int_0^\lambda d\lambda_1 f_1 (\lambda_1) I_\gamma (\omega_2,...,\omega_k; \lambda_1),
\end{align}
and, the 0-fold iterated integral satisfies $I_\gamma (;\lambda) =1$ ~\cite{Brown:2013qva}. 
These iterated integrals obey the shuffle algebra.
Very often Feynman integrals give rise to one-forms $\omega$'s which are of the form dlog~$p_i(x_1,..,x_m)$ where $x_i,...,x_n$ are the coordinates and $p_i(x_1,...,x_m)$ is an algebraic function, also known as a letter. Multiple polylogarithms (MPLs) are special cases of iterated integrals when $p_i$ is rational and linear-reducible in the variables $x_i$. For more details regarding the conditions of expressing iterated integrals as MPLs, one may refer to \cite{Duhr:2020gdd}. For non-rational $p_i$ (algebraic letters with square roots), there is no general algorithm to express the iterated integrals in terms of MPLs. Nevertheless, in physics, many examples of iterated integrals with dlog-forms with non-rational arguments appear, which can also be evaluated in terms of MPLs \cite{Duhr:2021fhk}.  
In order to present the analytic results in terms of iterated integrals over algebraic dlog one-forms, we need to bring the epsilon-factorized differential equation system to a dlog-form.
In the dlog-form, the entries of the matrix $\tilde{A}$ are $\mathbb{Q}$-linear combination of dlog one-forms so that 
\begin{equation}
\tilde{A}_{ij}= \sum_{k=1}^{31} \tilde{c}_{ijk} d\; \text{ln} (p_k(x,y,z)), \quad \tilde{c}_{ijk} \in \mathbb{Q}. 
\end{equation}
From the matrix $\tilde{A}$, a set of rational letters can be trivially obtained for the rational one-forms. For the one-forms with a dependence on $r$, getting a dlog one-form and therefore the list of non-rational letters, is not a trivial exercise.  
To construct the non-rational letters, we implement the algorithm from~\cite{Heller:2019gkq,Zoia:2021zmb} explained in the following:
\begin{enumerate}
\item We find the list of all the rational letters and the list of square roots. In our case, we have only one square root $r$ with a degree-8 polynomial.
\item
We construct all possible monomials upto degree 8 using the rational letters including  
$r^2$. Further, we identify those monomials which can be factorized as $(q_a+r)(q_a-r)$. This factorization gives the list of $q_a$'s. 

\item Using these $q_a$'s, we construct ansätze $l_a$ for dlog one-forms as
\begin{align}
l_a= \frac{q_a+r}{q_a -r}.
\end{align}
\item Using these ansätze, we fit all the non-rational one-forms which gives us a minimal set of dlog one-forms $p_a$.
\end{enumerate}

The complete alphabet with 31 letters $p_a$ that we obtain after the above steps is given by
\begin{align*}{\label{eqn:one_forms}}
&p_1 = x,\\
&p_2 = x-1,\\
&p_3 = x+1,\\
&p_4 = y,\\
&p_5 = y-1,\\
&p_6 = y+1,\\
&p_7 = z,\\
&p_8 = z-1,\\
&p_9 = z+1,\\
&p_{10}=x^2 y-x \left(y^2+1\right) z+y z^2,\\
&p_{11}=-y \left(x^2 z^2+1\right)+x y^2 z+x z,\\
&p_{12}=-x^2 y^2 z+x y \left(z^2+1\right)-z,\\
&p_{13}=x^2 y z-x \left(y^2 z^2+1\right)+y z,\\
&p_{14}=x y^2 z-x y z+x z-y z^2+y z-y,\\
&p_{15}=x^2 z-x \left(y z^2-y z+y+z\right)+z,\\
&p_{16}=x^2 y z-x \left((y-1) z+z^2+1\right)+y z,\\
&p_{17}=x^2 y z-x \left(y^2+y (z-1)+1\right)+y z,\\
&p_{18}=-y \left(x^2+x (z-1)+1\right)+x y^2 z+x z,\\
&p_{19} = \frac{-r +x^2 y z-x y^2 z-x y z^2-x y+x z+y z}{r +x^2 y z-x y^2 z-x y z^2-x y+x z+y z},\\ 
&p_{20} = \frac{-r +x^2 y z-2 x^2 y+x y^2 z-x y z^2+x y+x z-y z}{r +x^2 y z-2 x^2 y+x y^2 z-x y z^2+x y+x z-y z},\\
&p_{21} = \frac{-r +2 x^2 y^2 z-x^2 y z-x y^2 z-x y z^2-x y+x z+y z}{r +2 x^2 y^2 z-x^2 y z-x y^2 z-x y z^2-x y+x z+y z },\\
&p_{22} = \frac{-r +x^2 y z-2 x y^2 z^2+x y^2 z+x y z^2-x y-x z+y z}{r +x^2 y z-2 x y^2 z^2+x y^2 z+x y z^2-x y-x z+y z },\\
&p_{23} = \frac{-r +2 x^2 y z^2-x^2 y z-x y^2 z-x y z^2+x y-x z+y z}{r +2 x^2 y z^2-x^2 y z-x y^2 z-x y z^2+x y-x z+y z },\\
&p_{24} = \frac{-r +x^2 y z-2 x y^2 z^2+3 x y^2 z-2 x y^2+x y z^2-2 x y z+x y-x z+y z}{r +x^2 y z-2 x y^2 z^2+3 x y^2 z-2 x y^2+x y z^2-2 x y z+x y-x z+y z },\\
&p_{25} = \frac{-r +2 x^2 y^3 z-2 x^2 y^2 z+x^2 y z-2 x y^2 z^2+x y^2 z-2 x y^2+x y z^2+x y-x z+y z}{r +2 x^2 y^3 z-2 x^2 y^2 z+x^2 y z-2 x y^2 z^2+x y^2 z-2 x y^2+x y z^2+x y-x z+y z },\\
&p_{26} = \frac{-r +2 x^2 y z^3-2 x^2 y z^2+x^2 y z-2 x y^2 z^2+x y^2 z+x y z^2-x y-2 x z^2+x z+y z}{r +2 x^2 y z^3-2 x^2 y z^2+x^2 y z-2 x y^2 z^2+x y^2 z+x y z^2-x y-2 x z^2+x z+y z},\\
&p_{27} = \frac{-r +2 x^3 y^2 z-2 x^2 y^2 z-2 x^2 y z^2+x^2 y z-2 x^2 y+x y^2 z+x y z^2+x y+x z-y z}{r +2 x^3 y^2 z-2 x^2 y^2 z-2 x^2 y z^2+x^2 y z-2 x^2 y+x y^2 z+x y z^2+x y+x z-y z},\\
&p_{28} = \frac{-r +2 x^3 y-2 x^2 y^2 z+x^2 y z-2 x^2 y-2 x^2 z+x y^2 z+x y z^2+x y+x z-y z}{r +2 x^3 y-2 x^2 y^2 z+x^2 y z-2 x^2 y-2 x^2 z+x y^2 z+x y z^2+x y+x z-y z },\\
&p_{29} = \frac{-r +2 x^3 y z^2-2 x^2 y^2 z-2 x^2 y z^2+x^2 y z-2 x^2 z+x y^2 z+x y z^2+x y+x z-y z}{r +2 x^3 y z^2-2 x^2 y^2 z-2 x^2 y z^2+x^2 y z-2 x^2 z+x y^2 z+x y z^2+x y+x z-y z},\\
&p_{30} = (x y - z) (-y + x z) (x - y z) (-1 + x y z),\\
&p_{31}=r.
\end{align*}
We also find that the matrix $\tilde{A}$ in equation \ref{eqn:epsilon-form} contains only 31 $\mathbb{Q}$-independent linear combinations of dlog {one-forms.} Out of these 31 independent one-forms, 16 are rational in the variables $x, y$ and $z$ whereas 15 contain the square root $r$.

In recent times, iterated integral representation with dlog one-forms has been shown to be very efficient for phenomenological applications, see for example~\cite{Chicherin:2021dyp,Hartanto:2022qhh}. Iterated integrals with logarithmic one-forms have a clear branch cut structure, and the results expressed in terms of these functions have a compact analytic form. For a numerical evaluation of these functions, we can use a local series expansion of the one-forms combined with path-decomposition property satisfied by iterated integrals, see for example~\cite{Badger:2021owl}, to integrate from one phase-space point to another. Logarithmic one-forms can be series expanded, and since they have a power-log expansion a numerical evaluation is easy to implement. A representation in terms of iterated integrals is even useful for cases where public tools cannot be used due to an absence of power-log representation of the one-forms~\cite{Chaubey:2021ret}.

\section{Results and checks}
\label{sec:analytic_results}

Regarding the results presented in this article, we numerically evaluate the iterated integrals using an in-house implementation in {\tt Mathematica}. For the reader's convenience, we explain the steps that can be taken to do the numerical evaluation of iterated integrals in the following: 
\begin{enumerate}
\item we parametrize the one-forms on a path,
\item we series expand the one-dimensional one-forms around a point on the path,
\item we perform the iterated integration of the expanded one-forms.
\end{enumerate}
We might need multiple path segments if the phase-space point is far from the boundary point. In these cases, we perform the steps mentioned above on each path segment and use the path-decomposition formula~\cite{Brown:2013qva,Chaubey:2021ret} to obtain the final numerical result.
For phenomenological applications, the numerical evaluation of these functions can also be combined with several publicly-available tools for fast evaluation. For example, one can {set up} a differential system just for the iterated integrals and evaluate them with generalized power series expansions~\cite{Moriello:2019yhu} using tools like DiffExp \citep{Hidding:2020ytt} or parametrize the one-forms on a path and use GiNaC~\cite{VOLLINGA2005177}. 

As explained in the previous section, the result of master integrals can be written as Taylor expansion in $\epsilon$. Since our 
results are applicable to calculations involving two-loop integrals, we calculate master integrals only up to ${\cal O}(\epsilon^4)$.  
We provide results for all the 41 master integrals of canonical basis in terms of the iterated integrals with the dlog one-forms
in the supplementary material attached to this paper. We want to emphasize once again that the choice of the coordinate system given in equation (\ref{eqn:xyz}) allows us to write the iterated integrals appearing in $J_2-J_7$, $J_9, J_{10}$, $J_{12}-J_{17}$, $J_{19}$ in terms of MPLs.  

The complete results for the master integrals depend on the boundary terms.  
To obtain the analytic form of these boundary constants, we first evaluate the integrals in a regular limit~\cite{Chaubey:2019lum, Liu:2022chg}. We then match these values against the functional part of the results by first evaluating the iterated integrals up to many (100) digits and then using PSLQ~\cite{article} to extract the analytic constants.
The analytic expressions of all the boundary constants $B_i=(J_i)_{|(x=0,y=0,z=0)}$ in equation 
\ref{eqn:expandedJ} upto ${\mathcal O}\left(\epsilon^4\right)$ are as follows:
\begin{align*}
B_{1}  &= 
 1 + \zeta_2 \epsilon^2 - \frac{2}{3} \zeta_3 \epsilon^3 + \frac{7}{4} \zeta_4 \epsilon^4,\\
B_{2} &= 
 2 \zeta_2 \epsilon^2 + 4 \zeta_3 \epsilon^3 + \frac{19}{2} \zeta_4 \epsilon^4,\\
B_{3} &=
- 10 \zeta_4 \epsilon^4, \\
      B_{4} &=
 -\zeta_2 \epsilon^2 - 11 \zeta_3 \epsilon^3 - \frac{29}{2} \zeta_4 \epsilon^4,\\ B_{5} &=
 6 \zeta_3 \epsilon^3 + \frac{13}{2} \zeta_4 \epsilon^4, \\ 
B_{6} &=
 2 \zeta_2 \epsilon^2 + 4 \zeta_3 \epsilon^3 + \frac{19}{2} \zeta_4 \epsilon^4, \\ B_{7} &=
 2 \zeta_2 \epsilon^2 + 4 \zeta_3 \epsilon^3 + \frac{19}{2} \zeta_4 \epsilon^4,\\ 
 B_{8} &= 
 6 \iu\; \text{Im}\;  \Li_2\big(\frac{1}{2} (-1+\iu \sqrt{3})\big) \epsilon^2 - 
 \frac{\iu}{2}  \bigg(16\; \text{Im}\;\Li_3\big(\frac{\iu}{\sqrt{3}}\big) - 
   23\; \text{Im}\;\Li_3\big(\frac{1}{2}(-1+\iu\sqrt{3})\big) \\
   &+ 
   4\;  \text{Im}\;\mathrm{Li}_3\big(\frac{1}{2}(3 -\iu \sqrt{3})\big) \bigg)\epsilon^3 -
 \frac{\iu}{18} \bigg(92 \zeta_2 \;  \text{Im}\;\mathrm{Li}_2\big( \frac{1}{2} (-1 + \iu\sqrt{3})\big) + 
   92 \zeta_2 \;  \text{Im}\;\mathrm{Li}_2\big( \frac{1}{2}(3 -\iu \sqrt{3})\big) \\
   &- 
   288\;  \text{Im}\;\mathrm{Li}_4\big(\frac{\iu}{\sqrt{3}}\big) +
   207\;  \text{Im}\; \mathrm{Li}_4\big( \frac{1}{2}(-1 +\iu \sqrt{3})\big) +
   72\;  \text{Im}\;  \mathrm{Li}_4\big( \frac{1}{2} (3 - \iu\sqrt{3})\big)\bigg)
   \epsilon^4,\\
   B_{9}&= 
  - 10 \zeta_4\epsilon^4, \\ 
   B_{10} &=
  - 10 \zeta_4 \epsilon^4, \\  
  B_{11} &=
  12 \iu \zeta_2 \;\text{Im}\;  \mathrm{Li}_2\big(\frac{1}{2}(-1+\iu\sqrt{3})\big)\epsilon^4, \\
  B_{12} &= 
 - \zeta_2 \epsilon^2 - 11 \zeta_3 \epsilon^3 - \frac{29}{2} \zeta_4 \epsilon^4, \\
 B_{13} &= 
  6 \zeta_3 \epsilon^3 + \frac{13}{2} \zeta_4 \epsilon^4, \\
  B_{14} &=
 -\zeta_2 \epsilon^2 - 11 \zeta_3 \epsilon^3 - \frac{29}{2} \zeta_4 \epsilon^4, \\
 B_{15} &=
  6 \zeta_3 \epsilon^3 + \frac{13}{2} \zeta_4 \epsilon^4, \\
  B_{16}&=
 -10 \zeta_4 \epsilon^4, \\
 B_{17} &=
 -10 \zeta_4 \epsilon^4, \\
 B_{18} &=
12 \iu \zeta_2 \; \text{Im}\;\mathrm{Li}_2\big(\frac{1}{2}(-1 + \iu\sqrt{3})\big)
   \epsilon^4, \\
 B_{19} &=
 -10 \zeta_4 \epsilon^4, \\
 B_{20} &= 
12 \iu \zeta_2 \; \text{Im}\;\mathrm{Li}_2\big(\frac{1}{2}(-1 + \iu\sqrt{3})\big)
   \epsilon^4,\\
 B_{21} &=
-6 \iu\; \text{Im}\; \mathrm{Li}_2\big(\frac{1}{2}  (-1+ \iu\sqrt{3})\big) \epsilon^2 -\frac{\iu}{2} \bigg(-16\; \text{Im}\;  \mathrm{Li}_3\big( \frac{\iu}{\sqrt{3}}\big) + 
   23 \; \text{Im}\; \mathrm{Li}_3\big(\frac{1}{2} (-1 +\iu \sqrt{3})\big) \\
   &- 
   4\; \text{Im}\;  \mathrm{Li}_3\big(\frac{1}{2} (3 -\iu \sqrt{3})\big)\bigg) \epsilon^3  +\frac{\iu}{9} \bigg(100  \zeta_2 \; \text{Im}\;\mathrm{Li}_2\big(\frac{1}{2} (-1 + \iu\sqrt{3})\big) + 
  46 \zeta_2\; \text{Im}\; \mathrm{Li}_2\big(\frac{1}{2} (3 -\iu \sqrt{3})\big) \\
  &- 
   144 \; \text{Im}\; \mathrm{Li}_4\big(\frac{\iu}{\sqrt{3}}\big)- 
   18\; \text{Im}\;\mathrm{Li}_4\big( \frac{1}{2} (-1 + \iu\sqrt{3})\big) + 
   36 \; \text{Im}\;\mathrm{Li}_4\big(\frac{1}{2}(3 -\iu \sqrt{3})\big)\bigg) \epsilon^4, \\
 B_{22} &= 
\frac{\iu}{2}\bigg(12\zeta_2 \; \text{Im}\;\mathrm{Li}_2\big(\frac{1}{2}(-1 +\iu \sqrt{3})\big) - 
   27\; \text{Im}\; \mathrm{Li}_4\big( \frac{1}{2}(-1 + \iu\sqrt{3})\big)\bigg)\epsilon^4, \\
 B_{23} &=
\zeta_2 \epsilon^2 + 10 \zeta_3 \epsilon^3 + \frac{1}{4} \bigg(47 \zeta_4 - 36 \; \text{Im}\;\mathrm{Li}_2\big(\frac{1}{2}(-1 + \iu \sqrt{3})\big)^2\bigg) \epsilon^4, \\   
  B_{24} &=
-6 \iu\; \text{Im}\; \mathrm{Li}_2\big(\frac{1}{2}  (-1+ \iu\sqrt{3})\big) \epsilon^2 -\frac{\iu}{2} \bigg(-16\; \text{Im}\;  \mathrm{Li}_3\big( \frac{\iu}{\sqrt{3}}\big) + 
   23 \; \text{Im}\; \mathrm{Li}_3\big(\frac{1}{2} (-1 +\iu \sqrt{3})\big) \\
   &- 
   4\; \text{Im}\;  \mathrm{Li}_3\big(\frac{1}{2} (3 -\iu \sqrt{3})\big)\bigg) \epsilon^3  +\frac{\iu}{9} \bigg(100  \zeta_2 \; \text{Im}\;\mathrm{Li}_2\big(\frac{1}{2} (-1 + \iu\sqrt{3})\big) + 
  46 \zeta_2\; \text{Im}\; \mathrm{Li}_2\big(\frac{1}{2} (3 -\iu \sqrt{3})\big) \\
  &- 
   144 \; \text{Im}\; \mathrm{Li}_4\big(\frac{\iu}{\sqrt{3}}\big)- 
   18\; \text{Im}\;\mathrm{Li}_4\big( \frac{1}{2} (-1 + \iu\sqrt{3})\big) + 
   36 \; \text{Im}\;\mathrm{Li}_4\big(\frac{1}{2}(3 -\iu \sqrt{3})\big)\bigg) \epsilon^4, \\
 B_{25} &= 
\frac{\iu}{2}\bigg(12\zeta_2 \; \text{Im}\;\mathrm{Li}_2\big(\frac{1}{2}(-1 +\iu \sqrt{3})\big) - 
   27\; \text{Im}\; \mathrm{Li}_4\big( \frac{1}{2}(-1 + \iu\sqrt{3})\big)\bigg)\epsilon^4, \\
 B_{26} &=
-\zeta_2 \epsilon^2 - 10 \zeta_3 \epsilon^3 - \frac{1}{4} \bigg(47 \zeta_4 - 36 \; \text{Im}\;\mathrm{Li}_2\big(\frac{1}{2}(-1 + \iu\sqrt{3})\big)^2\bigg) \epsilon^4, \\   
 B_{27} &= 
-6 \iu\; \text{Im}\; \mathrm{Li}_2\big(\frac{1}{2}  (-1+ \iu\sqrt{3})\big) \epsilon^2 -\frac{\iu}{2} \bigg(-16\; \text{Im}\;  \mathrm{Li}_3\big( \frac{\iu}{\sqrt{3}}\big) + 
   23 \; \text{Im}\; \mathrm{Li}_3\big(\frac{1}{2} (-1 +\iu \sqrt{3})\big) \\
   &- 
   4\; \text{Im}\;  \mathrm{Li}_3\big(\frac{1}{2} (3 -\iu \sqrt{3})\big)\bigg) \epsilon^3  +\frac{\iu}{9} \bigg(100  \zeta_2 \; \text{Im}\;\mathrm{Li}_2\big(\frac{1}{2} (-1 + \iu\sqrt{3})\big) + 
  46 \zeta_2\; \text{Im}\; \mathrm{Li}_2\big(\frac{1}{2} (3 -\iu \sqrt{3})\big) \\
  &- 
   144 \; \text{Im}\; \mathrm{Li}_4\big(\frac{\iu}{\sqrt{3}}\big)- 
   18\; \text{Im}\;\mathrm{Li}_4\big( \frac{1}{2} (-1 + \iu\sqrt{3})\big) + 
   36 \; \text{Im}\;\mathrm{Li}_4\big(\frac{1}{2}(3 -\iu \sqrt{3})\big)\bigg) \epsilon^4, \\
 B_{28}&= 
 \frac{\iu}{2}\bigg(12\zeta_2 \; \text{Im}\;\mathrm{Li}_2\big(\frac{1}{2}(-1 +\iu \sqrt{3})\big) - 
   27\; \text{Im}\; \mathrm{Li}_4\big( \frac{1}{2}(-1 +\iu \sqrt{3})\big)\bigg)\epsilon^4, \\
 B_{29} &=
\zeta_2 \epsilon^2 + 10 \zeta_3 \epsilon^3 + \frac{1}{4} \bigg(47 \zeta_4 - 36 \; \text{Im}\;\mathrm{Li}_2\big(\frac{1}{2}(-1 + \iu\sqrt{3})\big)^2\bigg) \epsilon^4, \\    
 B_{30} &= 
 \frac{27\iu}{2} \; \text{Im}\; \mathrm{Li}_4\big(\frac{1}{2} (-1 + \iu\sqrt{3})\big) \epsilon^4, \\ 
  B_{31}&= 
 -\frac{\iu}{2}\bigg(-12 \zeta_2 \; \text{Im}\;\mathrm{Li}_2\big( \frac{1}{2} (-1 + \iu\sqrt{3})\big) - 
   27 \; \text{Im}\;\mathrm{Li}_4\big( \frac{1}{2} (-1 + \iu\sqrt{3})\big)\bigg) \epsilon^4,\\
 B_{32} &=
  -\frac{\iu}{2}\bigg(-12 \zeta_2 \; \text{Im}\;\mathrm{Li}_2\big( \frac{1}{2} (-1 + \iu\sqrt{3})\big) - 
   27 \; \text{Im}\;\mathrm{Li}_4\big( \frac{1}{2} (-1 + \iu\sqrt{3})\big)\bigg) \epsilon^4, \\
 B_{33}&= 
 -2 \zeta_3 \epsilon^3 + \frac{1}{2} \bigg(- \zeta_4 - 36 \; \text{Im}\;\mathrm{Li}_2\big(\frac{1}{2}(-1 + \iu\sqrt{3})\big)^2\bigg) \epsilon^4, \\ 
 B_{34} &=
 \frac{27\iu}{2} \; \text{Im}\; \mathrm{Li}_4\big(\frac{1}{2} (-1 + \iu\sqrt{3})\big) \epsilon^4, \\ 
  B_{35}&= 
  -\frac{\iu}{2}\bigg(-12 \zeta_2 \; \text{Im}\;\mathrm{Li}_2\big( \frac{1}{2} (-1 + \iu\sqrt{3})\big) - 
   27 \; \text{Im}\;\mathrm{Li}_4\big( \frac{1}{2} (-1 + \iu\sqrt{3})\big)\bigg) \epsilon^4, \\
 B_{36} &=
  -\frac{\iu}{2}\bigg(-12 \zeta_2 \; \text{Im}\;\mathrm{Li}_2\big( \frac{1}{2} (-1 + \iu\sqrt{3})\big) - 
   27 \; \text{Im}\;\mathrm{Li}_4\big( \frac{1}{2} (-1 + \iu\sqrt{3})\big)\bigg) \epsilon^4, \\
 B_{37}&=
 -2 \zeta_3 \epsilon^3 + \frac{1}{2} \bigg(- \zeta_4 - 36 \; \text{Im}\;\mathrm{Li}_2\big(\frac{1}{2}(-1 +\iu \sqrt{3})\big)^2\bigg) \epsilon^4, \\ 
 B_{38} &= 
 \frac{27\iu}{2} \; \text{Im}\; \mathrm{Li}_4\big(\frac{1}{2} (-1 + \iu\sqrt{3})\big) \epsilon^4, \\ 
  B_{39} &=
  -\frac{\iu}{2}\bigg(-12 \zeta_2 \; \text{Im}\;\mathrm{Li}_2\big( \frac{1}{2} (-1 + \iu\sqrt{3})\big) - 
   27 \; \text{Im}\;\mathrm{Li}_4\big( \frac{1}{2} (-1 + \iu\sqrt{3})\big)\bigg) \epsilon^4, \\
 B_{40}&= 
  -\frac{\iu}{2}\bigg(-12 \zeta_2 \; \text{Im}\;\mathrm{Li}_2\big( \frac{1}{2} (-1 + \iu\sqrt{3})\big) - 
   27 \; \text{Im}\;\mathrm{Li}_4\big( \frac{1}{2} (-1 + \iu\sqrt{3})\big)\bigg) \epsilon^4, \\
 B_{41} &=
 -2 \zeta_3 \epsilon^3 +\frac{1}{2} \bigg(- \zeta_4 - 36 \; \text{Im}\;\mathrm{Li}_2\big(\frac{1}{2}(-1 + \iu \sqrt{3})\big)^2\bigg) \epsilon^4.\\
\end{align*}

To verify the correctness of our results, we numerically evaluate the master integrals at multiple phase-space points. For this, we parametrize the one-forms on a path, use a series expansion of the integrands around ($x=0, y=0, z=0$), and integrate the iterated integrals at some phase-space point. We then match the results against the numerical values obtained from pySecdec \citep{Borowka_2018} and AMFlow \cite{Liu:2022chg} at the same phase-space point and find good agreement ($\sim$ 100 digits). We present the numerical results of one of the master integrals from the 
top sectors $J_{41}$ up to $\mathcal{O}$($\epsilon^4$) evaluated at point $(x,y,z)$ = (0.5, 0.5, 0.5) up to 90 decimal places in table \ref{tab:numerical_results}. 
\begin{table}[!htp]
\begin{center}
\footnotesize\addtolength{\tabcolsep}{-5pt}
\begin{tabular}{|c|c|}
\hline
$\mathcal{O}$  &  $J_{41}$ \\
\hline
 $\epsilon^0$ & 0 \\
\hline
$\epsilon^1$ & 0\\
\hline
 $\epsilon^2$& 0.480453013918201424667102526326664971730552951594545586866864133623665382259834472199948263   \\
\hline
$\epsilon^3$ & 0.091897357209531231888638645571102756639040958784768669397674639926722578733737539395677303 \\
\hline
$\epsilon^4$ & 0.778715275596109429538387498220501297387117166408046400975462356038382264999189687267373506\\
\hline
\end{tabular}
\end{center}
\caption{Numerical values of $J_{41}$ evaluated at $(x,y,z)$ = (0.5, 0.5, 0.5) for first five orders of $\epsilon$ using analytic results.
}
\label{tab:numerical_results}
\end{table}
\section{Conclusions}
We have presented the analytic results for the master integrals that contribute to the mixed electroweak-QCD corrections in 
$H \rightarrow Z Z^{*}$ decay. We obtain a canonical form of the differential equations for all the master integrals. Despite the simultaneous non-rationalizability of all the square roots that appear in the differential equations, we construct ansätze to find dlog-forms having algebraic dependence on the square roots. Our results are obtained in terms of Chen's iterated integrals order-by-order in $\epsilon$, the parameter of dimensional regularization. Numerical evaluations of the result have been performed, and checks have been carried out using publicly available tools. The results presented here are straightforwardly applicable for computing {the} partial decay width of 
{$H \to ZZ^* \to 4\ell$} at ${\cal O}(\alpha \alpha_s)$ accuracy and, in any production or decay process involving two-loop three-point Feynman diagrams with four mass scales. Our analytic results can be used in event generation codes for Higgs production and decay to carry out phenomenological studies and data analysis for future collider experiments such as $e^+e^-$, $\mu^+\mu^-$ and FCC-hh colliders.

\label{sec:conclusion}

\section*{Acknowledgements}
EC receives funding from the European Union’s Horizon 2020 research and innovation programmes
\textit{High precision multi-jet dynamics at the LHC} (grant agreement No {772099}). EC would like to thank Vasily Sotnikov for useful discussions. MK would like to acknowledge financial support from IISER Mohali for
this work.

\newpage

\bibliographystyle{JHEP}
\bibliography{bibrefer} 
\end{document}